\documentclass[12pt,a4paper]{article}

\usepackage{setspace,graphics}
\usepackage[dvips]{epsfig} 
\usepackage{a4,amssymb,epsfig,array,cite}
\usepackage[hypertex]{hyperref}

\newcounter{multieqs}

\newcommand{\be}{\begin{equation}}
\newcommand{\ee}{\end{equation}}
\newcommand{\eq}[1]{(\ref{#1})}

\def\nn{\nonumber}
\def\bea{\begin{eqnarray}}
\def\eea{\end{eqnarray}}

\def\beqa{\begin{eqnarray}} 
\def\eeqa{\end{eqnarray}} 
\def\beq{\begin{equation}} 
\def\eeq{\end{equation}}

\def\a{\alpha}          
\def\b{\beta}           
  \def\C{\Gamma}  
\def\d{\delta}    
\def\e{\epsilon}

 \def\L{\Lambda} \def\la{\lambda}

\def\s{\sigma}

  \def\cC{{\cal C}}
\def\cD{{\cal D}}  
  
\def\cJ{{\cal J}}  \def\cL{{\cal L}}
  \def\cO{{\cal O}}
\def\cP{{\cal P}}

\def\mg{\mathfrak{g}}
\def\mh{\mathfrak{h}}
\def\mr{\mathfrak{r}}
\def\mt{\mathfrak{t}}

\def\R{{\mathbb R}}
\def\C{{\mathbb C}}

\def\Z{{\mathbb Z}}
\def\one{\mbox{1 \kern-.59em {\rm l}}}

\def\bit{\begin{itemize}}
\def\eit{\end{itemize}}

\def\({\left(}
\def\){\right)}

\def\tens{\otimes}

\def\rep{representation }
\def\reps{representations }

\def\d{\delta}

 \def\del{\partial}

\def\uno{\mbox{1 \kern-.59em {\rm l}}}

\def\one{1\!\!1\,\,}

\def\bcomment#1{}

\renewcommand{\title}[1]{\vspace{10mm}\noindent{\Large{\bf #1}}\vspace{8mm}}
\newcommand{\authors}[1]{\noindent{\large #1}\vspace{5mm}}
\newcommand{\address}[1]{{\itshape #1\vspace{2mm}}}

\begin{document}

\begin{titlepage}

\begin{flushright}
LMU-TPW 04-05 \\
UWTHPh-2004-16\\
\end{flushright}

\begin{center}
  
\title{Finite Gauge Theory on Fuzzy $\C P^2$ }

\authors{Harald {\sc Grosse}$^1$ and Harold {\sc Steinacker}$^2$}

\address{$^{1}$\,Institut f\"ur Theoretische Physik, Universit\"at Wien\\
Boltzmanngasse 5, A-1090 Wien, Austria}

\address{$^{2}$\,Institut f\"ur theoretische Physik \\
Ludwig--Maximilians--Universit\"at M\"unchen \\
Theresienstr.\ 37, D-80333 M\"unchen, Germany}

\footnotetext[1]{harald.grosse@univie.ac.at}
\footnotetext[2]{harold.steinacker@physik.uni-muenchen.de}

\vskip 2cm

\textbf{Abstract}

\vskip 3mm

\begin{minipage}{14cm}%
We give a non-perturbative definition of $U(n)$ gauge theory on 
fuzzy $\C P^2$ as a multi-matrix model. 
The degrees of freedom are
8 hermitian matrices of finite size, 4 of which are
tangential gauge fields and 4 are auxiliary variables.
The model depends on a noncommutativity parameter $\frac 1N$, 
and reduces to the usual $U(n)$ Yang-Mills
action on the 4-dimensional classical
$\C P^2$ in the limit $N \to \infty$. 
We explicitly find the monopole solutions, and also 
certain $U(2)$ instanton solutions for finite $N$. 
The quantization of the model is defined in terms of a path integral,
which is manifestly finite. An alternative formulation with
constraints is also given, and
a scaling limit as $\R^4_\theta$ is discussed.
\end{minipage}

\end{center}

\end{titlepage}

\setcounter{page}0
\thispagestyle{empty}
%\newpage

%%%%%%%%% table of content %%%%%%%
\begin{spacing}{.3}
{
\noindent\rule\textwidth{.1pt}            % THEN MAKE TOC...
   \tableofcontents
\vspace{.6cm}
\noindent\rule\textwidth{.1pt}
}
\end{spacing}

%%%%%%%%%%%%%% ordinary document (end) ####################################
\section{Introduction}

Fuzzy spaces are a nice class of noncommutative spaces, 
which  admit only finitely many degrees of freedom,
but are compatible with the classical symmetries.
This means that field theory on fuzzy spaces is regularized,
but compatible with a geometrical symmetry group unlike
lattice field theory. A large 
family of such spaces is given by the quantization of (co)adjoint
orbits $\cO$ of a Lie group in terms of certain finite matrix
algebras $\cO_N$. They are labeled by a 
a noncommutativity parameter $\frac 1N$, and the classical space
is recovered in the large $N$ limit.
The simplest example is the fuzzy sphere $S^2_N$, which has been
studied in great detail; see e.g.
\cite{madore,Madore1991,
grosse1,klimcik,baez,rupp,watamura,iso,fuzzyloop,s2inst,matrixsphere,ydri-perturb}
and references therein. 
The purpose of 
this paper is to find a useful formulation of gauge theory on 
the 4-dimensional fuzzy space $\C P^2_N$, and to study some of its properties 
including topologically nontrivial solutions. 

The most
obvious 4-dimensional fuzzy spaces are
$S^2_N \times S^2_N$ and $\C P^2_N$. While the former is 
technically easier to handle,  $\C P^2_N$ 
(see e.g. \cite{nair-cp2,stroh,bala,cpn-bala,azuma})
has an 8-dimensional
symmetry group $SU(3)$, which is larger 
than that of $S^2_N \times S^2_N$. This
leads to the hope that more can be done on $\C P^2_N$.
We propose a definition of gauge theory on $\C P^2_N$ in terms of 
certain multi-matrix models, generalizing the approach of
\cite{matrixsphere}. Our basic requirement is that
it should reduce 
to the usual $U(n)$ Yang-Mills gauge theory on classical $\C P^2$
in the commutative limit, but it should also be simple and
promise advantages over the commutative case. 

It is a well-known and fascinating fact that gauge theory on
noncommutative spaces can be formulated in terms of multi-matrix
models. Such matrix models also arise 
in string theory, e.g. the IKKT matrix model \cite{ikkt}
and effective actions for certain D-branes \cite{ARS}.
This leads to a picture where the space (a ``brane'')
arises dynamically as solution  of such an action, and can be
interpreted as submanifold of a 
higher-dimensional space. The gauge fields arise as fluctuations of
the tangential coordinates, and 
the transversal coordinates become 
additional scalar fields on the brane. While the matrices are
usually infinite-dimensional, 
they are finite-dimensional on fuzzy spaces.

For the fuzzy sphere, 
a formulation of gauge theory as matrix model was first
given in \cite{watamura}, and a model which reduces to the classical 
Yang-Mills theory on $S^2$ in the large $N$ limit was studied 
in detail and quantized in \cite{matrixsphere}.
The formulation  as matrix model 
has at least 2 notable features, which are not present in other 
formulations or in the classical case: First, it
leads to a very simple picture of nontrivial gauge sectors
such as monopoles, which arise as nontrivial 
solutions in the matrix configuration space.
This was noted in \cite{nair} and fully explored in
\cite{matrixsphere} for the fuzzy sphere; see also \cite{valtancoli}
for related work.
The concepts of fiber bundles are not required but arise
automatically, in an intrinsically noncommutative way. 
Second, it allows a nonperturbative quantization 
in terms of a finite ``path'' integral,
which in the case of $U(n)$ Yang-Mills on $S^2_N$ 
can be carried out explicitly in the large $N$ limit \cite{matrixsphere}.
We want to see if these features can be extended to $\C P^2_N$.
It turns out that we can indeed find monopole and (generalized)
instanton solutions  on $\C P^2_N$, generalizing 
the approach of \cite{matrixsphere}.

Because the paper is rather long, we briefly outline the main steps here.
We start with a detailed review of classical 
and fuzzy $\C P^2$ in Sections \ref{sec:class-geom} and \ref{sec:fuzzyCP2},
including some aspects of (co)homology and differential forms. We also
give a useful new link between the fuzzy and the classical
differential calculus \eq{dxi}.
Furthermore, we develop some tools
in order to reformulate the usual tangential tensor fields in terms of 
$su(3)$ tensor fields, which arise through the embedding 
$\C P^2 \hookrightarrow  \R^8  \cong su(3)$.
These tools are certain linear and nonlinear $su(3)$-equivariant 
tensor maps, which are 
discussed in Section \ref{sec:tensor-maps} for the commutative case, 
and in Section \ref{sec:nc-constraints}
in the noncommutative case. 
We also show in Section \ref{sec:plane}
how the canonical quantum plane $\R^4_\theta$ can be 
obtained from fuzzy $\C P^2$ in a scaling limit.

Using this background and following \cite{matrixsphere},
we propose the following action  
for $U(n)$ gauge theory on $\C P^2_N$ in Section \ref{sec:YM}:
\beq
S = \frac 1{g}\; \int tr F_{ab} F_{ab} + S_D.
\eeq
Here $F_{ab} = i[C_a,C_b] + \frac 12 f_{abc} C_c$ is the field strength, and
$S_D$ is a Casimir-type constraint term \eq{S_cons}. 
The basic variables are 
8 ``covariant coordinates'' $C_a$,  which describe the embedding
of $\C P^2 \hookrightarrow  \R^8$. Expanding them around the
``vacuum''
solution $C_a = \xi_a + A_a$ then leads to a gauge theory on $\C P^2_N$.
The crucial point is the addition of the
constraint term S$_D$, which is chosen such that
only configurations close to the vacuum solution $C_a = \xi_a$
which defines fuzzy $\C P^2$ are relevant. In other words, 
the constraint term $S_D$ stabilizes
the space $\C P^2 \subset \R^8$ by giving the transversal fluctuations
a large mass. The nontrivial part is to make sure that 4 physical,
tangential gauge fields survive, and that 
the standard 4-dimensional $U(n)$ Yang-Mills theory emerges in the
commutative limit.
The remaining 4 degrees of freedom become very massive scalars and decouple. 
This is somewhat subtle, and discussed in detail in Sections
\ref{sec:YM} and \ref{sec:decoupling}.
An alternative formulation of a gauge theory on $\C P^2_N$ is given in Section 
\ref{sec:constraints}, imposing a suitable constraint which admit 
only tangential fields.

A nontrivial test of the proposed models is to see whether they 
admit topologically
nontrivial solutions such as instantons. On classical 
$\C P^2$, there exist both $U(1)$ monopoles
as well as $SU(2)$ instantons. The $U(1)$ monopoles are labeled by an 
integer which corresponds
to the first Chern class, and have a selfdual field strength. Such monopoles
were constructed recently on fuzzy $\C P^2$ in  \cite{wataCPN} 
as projective modules. In our formulation,
it is quite straightforward to recover them as 
exact solutions of the equations of
motions, similar as in
\cite{matrixsphere}. In the commutative limit they become connections on the
monopole bundles of $\C P^2$, which we give explicitly.
In particular, we 
reproduce the results of \cite{wataCPN} without having to introduce additional
structure such as 
projective modules. In fact, all these monopole (and instanton) solutions 
arise in the same configuration space. 
This is a remarkable simplification over the commutative case,
and provides further support for this approach. 
Similarly, we find exact solutions 
for the nonabelian $U(2)$ case in Section \ref{sec:instantons}, 
which in the commutative limit describe certain nontrivial
rank 2 ``instantons''  on $\C P^2$. The classical bundle structure
is clarified in Section \ref{sec:class-bundle}, and 
the $U(2)$ connection is computed explicitly in Appendix E using the
Gelfand-Tsetlin basis for $su(3)$ (this takes up much of the space in the
appendix).
However, our purely group-theoretical 
ansatz only gives non-localized instantons, whose
fields strength is essentially constant. Finding localized instantons 
and their moduli space \cite{buchdahl} 
remains an interesting challenge. In particular, our
``instantons'' contain a nontrivial $U(1)$ component, and are neither
selfdual nor anti-selfdual. The $U(1)$ seems to be related with
the spin${}^c$ structure on $\C P^2$.

The quantization of this gauge theory is straightforward in principle,
in terms of a ``path integral'' which is convergent. As
opposed to the 2-dimensional case \cite{matrixsphere},
it can no longer be performed analytically. However, we point out an
interesting 2-matrix model \eq{S-VVp} which is in the class of the models 
discussed above, and which might be accessible to similar analytical
studies.

The presentation in this paper is detailed rather than 
short, but the
mathematical formalism is kept at a minimum.
We try to motivate the various choices made, discuss
alternatives, and explain how 
we arrive at our models.  This is sometimes done 
at the expense of space, but we hope that the amount of
results and details justify the length of this paper.

\section{Classical $\C P^2$}
\label{sec:class-geom}

For our purpose, the most useful description of $\C P^2$ is 
as a (co)adjoint orbit in $su(3)$. In general, 
they have the form 
\beq
\cO(t) = \{g t g^{-1}; \quad g \in G\}
\label{conj-classes}
\eeq
for some $t \in \mg = T_eG$. 
These conjugacy classes are invariant under the adjoint action of
$G$. Then $\cO(t)$ can be viewed as a homogeneous space:
\beq
\cO(t) \cong G/K_t
\label{coset}
\eeq
where $K_t = \{g \in G:\; Ad_g(t) = 0\}$ is the stabilizer of $t$.
The dimension and the type of $\cO(t)$ depends only on $K_t$.
For $G = SU(n)$, we can assume that 
$t$ is a diagonal matrix. 
In particular, in order to obtain $\C P^2 = SU(3)/(SU(2) \times U(1))$
we choose 
\beq
t = \tau_8 =  \frac 1{\sqrt{3}}\left(\begin{array}{ccc} 2 & 0& 0\\
                             0 & -1 & 0 \\
                             0 & 0 & -1 \end{array}\right).
\eeq
Here $\tau_a$ are the ``conjugated'' 
Gell-mann matrices\footnote{the reason for using $\tau_a$ instead of the
standard Gell-mann matrices is simply that we would like to have a north pole 
rather than a south pole. The conventions here are
different from \cite{wataCPN}.} 
 of $su(3)$, which satisfy
\beqa
tr(\tau_a \tau_b) &=& 2 \d_{ab},\nn\\
\tau_a \tau_b &=& \frac 23 \d_{ab} 
  + \frac 12(i {f_{ab}}^c + {d_{ab}}^c) \tau_c.
\label{tau-algebra}
\eeqa
They are given explicitly in  Appendix A, along with
the tensors $f_{abc}$ and $d_{abc}$. One can use \eq{coset} to 
derive the decomposition of the space of functions on $\C P^2$ into
harmonics i.e. 
irreps under the adjoint action of $SU(3)$ \cite{stroh,wataCPN},
\beq
\cC^{\infty}(\C P^2) = \mathop{\oplus}_{p=0}^\infty V_{(p,p)}.
\label{CP2-harmonics}
\eeq
Here $V_{(n,m)}$ denotes the irrep of $su(3)$ with 
highest weight $n \L_1 + m\L_2$, and $\L_i$ are the fundamental weights
of $su(3)$.

 It is convenient to work with
the over-complete set of global coordinates defined by the embedding
$\C P^2 \subset \R^8$ in the Lie algebra $su(3) \cong \R^8$. 
We can then write any element $X \in \C P^2$ as
\beq
\C P^2 = \{X= x_a \tau_a = g^{-1} t  g; \;\;t = \tau_8,\; g \in SU(3)\}.
\label{cp2coordinate}
\eeq 
It is characterized by the characteristic (matrix) equation 
\beq
X X = \frac 1{\sqrt{3}}\; X + \frac 23
\label{char-class}
\eeq
which is easy to check for $X = \tau_8$.
In component notation, this takes the form \cite{bala}
\beqa
g^{ab} x_a x_b    &=& 1, \label{def1c} \\
d^{ab}_c x_a  x_b &=& \frac{2}{\sqrt{3}}\; x_c. \label{def3c}
\eeqa
This can be understood as follows: The matrix
\beq
P = \frac 1{\sqrt{3}} (X+ \frac 1{\sqrt{3}})
\eeq
satisfies
\beq
P^2 = P, \qquad Tr(P) =1
\label{projector-class}
\eeq
as a consequence of \eq{char-class}, hence $P$ is a projector of rank 1.
Such projectors are equivalent via $P = |z_i\rangle \langle z_i|$ 
to complex lines in $\C^3$, which leads to the more familiar
definition of $\C P^2$.
Sometimes a radius $R$ will be introduced by rescaling $x_a \to x_a R$.

\paragraph{Some geometry.}

Notice that the symmetry group $SU(3)$ contains both ``rotations'' as
well as ``translations''. The generators $L_a$
act on an element $X = x_a \tau_a \in \C P^2$ as
\beq
L_a X = [\tau_a,X] = x_b [\tau_a,\tau_b] = i {f_{abc}}\; x_b \tau_c.
\eeq
In terms of the coordinate functions on the embedding space $\R^8$,
this can be realized as differential operator
\beq
L_a = \frac i2 {f}_{abc}(x_b \del_c - x_c \del_b).
\eeq
Now we can identify the rotations: 
consider the {\em ``north pole''}
$$
X_{np} = \tau_8 = x_a \tau_a \quad \in \C P^2 
\qquad \mbox{with} \;\;x_a = \d_{a,8}.
$$ 
The rotation subgroup is
its stabilizer subalgebra $\mr \cong su(2) \times u(1) \subset su(3)$ 
generated by  the ``rotation'' generators
\beq
\mr = \{\tau_1,\tau_2,\tau_3,\tau_8\}
\eeq
resp. the corresponding\footnote{we will sometimes denote the 
indices $1,2,3,8$ with $\mr$, etc} $L_\mr$.
$\mr$ is clearly a subalgebra
of the Euclidean rotation algebra $so(4) = su(2)_L \times su(2)_R$.
The translations of $X_{np}$ are generated by
the ``translation generators'' 
\beq
\mt = \{\tau_4,\tau_5,\tau_6,\tau_7\}.
\eeq
$\mt$ is not a subalgebra of $su(3)$, but the following relations 
hold\footnote{notice that the symmetric product defined by the 
$d$ symbol respects the 
same subspace structure.}:
\beq
[\mr,\mr] \subset \mr, \qquad [\mr,\mt] = \mt, 
 \qquad [\mt,\mt] = \mr.
\label{reductive}
\eeq
It is instructive to work out these generators 
in terms of differential operators at the north pole $x_a = R
\d_{a,8}$
(introducing a radius $R$ of $\C P^2$):
\beq
L_\mt = i R  f_{\mt 8 \mt'}\del_{\mt'}
\eeq
Hence 
\beq
L_4 =-\sqrt{3} i R\del_5,\quad L_5 = \sqrt{3} iR\del_4, \quad 
L_6 = - \sqrt{3} iR\del_7,\quad L_7  = \sqrt{3} iR\del_6
\label{L-transl-id}
\eeq
indeed generate the 4 translations at the north pole.

There is another interesting subspace of $\C P^2$: the ``south
sphere'' $S^2_S$ \cite{wataCPN} or ``sphere at infinity''.
This is a nontrivial cycle of $\C P^2$ which will play an important role later.
Consider again the parametrization of $\C P^2$ in terms of 
$3 \times 3$ matrices $X = U^{-1} \tau_8 U$
introduced above. Using a suitable 
$U \in SU(3)$, we can put it into the form
\beq
X = \left(\begin{array}{ccc} -\frac 1{\sqrt{3}} & \vline & 0\\
                                   \hline 
                      0 &\vline & \frac 1{2\sqrt{3}} + x_i \sigma^i \end{array}\right)
  = -\frac 12 \tau_8 +  \sum_{a=1,2,3} x_a \tau_a.
\label{southsphere}
\eeq
This is the subspace of $\C P^2$ with the most negative value of 
$x_8 = -\frac 12$, where
$x_{4,5,6,7} =0$ and $x_1^2+x_2^2+x_3^2 = \frac 34$. 
Hence this is a sphere of radius $\frac{\sqrt{3}}2$.

It is also quite illuminating to write down explicitly \eq{def3c}
for $i=1,2,3,8$. Using \eq{d-explicit}  and \eq{def1c}, they can be written as
\beqa
\frac 1{\sqrt{3}}\; (x_1+i x_2) &=& \frac{-1}{2x_8+1}(x_4+ix_5)(x_6-ix_7), 
       \nn\\ 
\frac 2{\sqrt{3}}\; x_3 &=& \frac{1}{2x_8+1}(x_6^2+x_7^2-x_4^2-x_5^2), 
       \label{constraint-explicit}\\
(1-x_8)(1+2x_8) &=& \frac 32 (x_4^2+x_5^2+x_6^2+x_7^2) \nn.
\eeqa
This shows how the ``transversal'' variables $x_{\mr}$ are expressed
in terms of the tangential ones. The other equations of \eq{def3c} are
redundand except at the south sphere.
In particular, note that fixing a value for $x_8$ determines a 
3-sphere $S^3 \subset \R^4_\mt = \langle x_4,x_5,x_6,x_7 \rangle$.
Then the first 2 equations above determine a map 
$S^3 \to S^2$, 
which is precisely the Hopf map. This shows explicitly that 
$\C P^2$ is a compactification of $\R^4$ via a
sphere ``at infinity''.

\subsection{$SU(3)$ tensors and constraints}
\label{sec:tensor-maps}

In order to better understand the fuzzy case, it is useful to 
consider tensor fields on $\C P^2$ with indices in the adjoint of $su(3)$;
these will be denoted as ``tensors'' throughout this paper.
Their transformation under the symmetry group $SU(3)$ is defined 
in the obvious way. 
Examples are the above global ``coordinates'' $x_a$, the field strength
$F_{ab}(x)$ defined below etc. 
The best way to  think of them is as
pull-backs via the embedding map  $\C P^2 \subset \R^8$
of ordinary tensors which live on  $\R^8$.
The ``rank'' of such tensors will be the number of $su(3)$ indices.
In order to relate them to the usual tangential tensor fields, we introduce
some useful maps which separate the tangential from the transversal
components. See also \cite{bala,cpn-bala} for similar
considerations.

\subsubsection{Linear tensor maps}

For rank one tensors (one-forms) 
$X = X_a dx_a \cong X_a \tau_a$ in the above sense,
consider the map \cite{cpn-bala}
\beq
J(X):= - \frac i{\sqrt{3}}\; [x^a\tau_a, X].
\label{cJ}
\eeq
In component form, it is a linear map of tensors
\beq
J(X)_a  = \frac 1{\sqrt{3}}\; f_{abc} x_b X_c.
\eeq
It is easy to see (e.g. at the north pole) that
it takes the eigenvalue 0  on 
non-tangential fields, and rotates the tangential ones with 
$J^2 = -1$ on tangential fields.
$J$ corresponds to the complex structure on $\C P^2$ \cite{bala}.

Similarly, consider the map\footnote{A similar operator was introduced
in \cite{cpn-bala}}
\beq
D^{lin}(X)_a  =  \frac 1{\sqrt{3}}\;  d_{abc} x_b X_c - \frac 13 X_a
\eeq
for any tensor $X$.  At the north pole, it is 
\beq
D^{lin}(X)_a =  \frac 1{\sqrt{3}}\;   d_{a8c} X_c - \frac 13 X_a
\eeq
which has the eigenvalues $-1$ for $a=1,2,3$, 
and $+\frac 1{3}$ for $a=8$, and
$0$ for $a=4,5,6,7$.
Therefore the space of {\bf tangential tensors} is globally characterized by
\beq
\fbox{$D^{lin}(X)_a = 0$.}
\eeq
Finally, the map
\beq
P(X)_a:= x_a (x_b X_b)
\eeq
projects on the ``radial'' part of $X$, and has eigenvalues $0,1$.

Let us see explicitly how $D^{lin}$ decomposes a tensor $X_a$.
Using 
the identity \eq{dd-rewritten}, one finds
\beq
(D^{lin})^2(X)_a  =  -\frac 23 D^{lin}(X)_a +\frac 13 X_a + \frac 19 J^2 (X)_a 
\label{DDeq}
\eeq
Moreover, the identity 
\beq
\sum_e (f^{ade} d^{bce} + f^{bde}  d^{eca}  + f^{cde} d^{eab} ) =0
\label{fd-identity}
\eeq
implies 
\beq
[J,D^{lin}] =0,
\eeq
while $[P,D] = 0 = [P,J]$ is obvious. Similarly, contracting the identity
\beq
\sum_e f^{abe} f^{cde} = \frac 83 (\d^{ac} \d^{bd} - \d^{bc} \d^{ad}) 
 + \sum_e (d^{ace} d^{bde} - d^{ade} d^{bce})
\eeq
 with $x_d x_b X_c$ gives
\beq
- J^2(X)_a + 3(D^{lin})^2(X)_a = -  \frac 83 P(X)_a
  + 3 X_a  
\eeq
which  together with \eq{DDeq} implies
\beq
(D^{lin})^2(X)_a+D^{lin}(X)_a  =  \frac 49 P(X)_a
\label{D2J2}
\eeq
Now suppose we have a fixed point 
$D^{lin}(X)_a = 0$, i.e. $X_a$ is tangential. 
Then \eq{D2J2} implies
$P(X)  = 0$, 
i.e. 
\beq
x\cdot X =  0
\eeq
as it should, i.e. the ``radial'' component vanishes.
Further, suppose we have a fixed point 
$D^{lin}(X)_a = \frac 13  X_a$. 
Then \eq{D2J2} gives
$X_a =  P(X)_a$,
hence $X_a$ is purely radial,
\beq
X_a = x_a f(x).
\eeq
The maps $D^{lin}$ and $P$ can be used to project any tensor field to its
tangential components, by 
\beq
X \to X_{tang} = X - \frac 43 P(X) + D^{lin}(X).
\label{tang-proj-approx-class}
\eeq

\subsubsection{Nonlinear tensor maps}
\label{subsubsec:nonlinmaps}

We want to understand better the constraint \eq{def3c}. In particular,
we can show that \eq{def3c}
implies \eq{def1c}. To see this, 
define the following nonlinear tensor map
\beq
D^{nl}(X)_a =  d_{abc} X_b X_c -\frac 2{\sqrt{3}} X_a.
\eeq
Assume we have some ``eigenvector'' of $D^{nl}$, which we write as
\beq
D^{nl}(X)_a = (\a-\frac 2{\sqrt{3}}) X_a
\eeq
for some constant $\a$, i.e. $d_{abc} X_b X_c = \a\; X_a$.
By rewriting $d X (dXX)$ using \eq{dd-rewritten} 
and $f_{abc} X_b X_c= 0$, one finds
\beq
\a^2\; X_a =  \frac 43 (X\cdot X) X_a 
\label{DDnleq}
\eeq
hence
\beq
\a^2 = \frac 43 (X\cdot X).
\eeq
In particular, it follows that $X\cdot X$ is a constant, and

{\em \eq{def1c} is a consequence of \eq{def3c}}.

Therefore a tensor field $X_a(x)$ which satisfies 
$D^{nl}(X) = 0$ describes 
functions $X: \C P^2 \to \C P^2$, with normalization $X \cdot X =1$.
Furthermore, note that for infinitesimal variations $\d X$ we have
$\d D^{nl}(X) = 2 \sqrt{3}\; D^{lin}(\d X)$. This
means that the linear constraint $D^{lin}(\d X) =0$
describes tangential fields on $\C P^2$, in agreement with the
previous findings.

In the noncommutative case, we will see that a fuzzy version of $X_a$
satisfying (approximately) $d_{abc} X_b X_c = \frac 2{\sqrt{3}} X_a$
admits 4 tangential degrees of freedom, which are identified as gauge
fields. Hence gauge theory can be interpreted as a
theory of fluctuating (covariant, fuzzy) coordinates of $\C P^2$.

\subsection{Symplectic form and (anti-) selfduality}
\label{subsec:selfdual-class}

$\C P^2$ is a symplectic (K\"ahler) space. The symplectic (K\"ahler)
form is given by\footnote{to see that it is closed, 
note that $d\eta \propto f_{abc} dx_a dx_b dx_c =0$ 
on $\C P^2$ due to the explicit
form of $f_{abc}$}
\beq
\eta = \frac 1{2\sqrt{3}R} f_{abc} x_a dx_b dx_c
\label{eta-class}
\eeq
which is clearly invariant under $SU(3)$. Here $\eta$ is normalized such that 
$\langle \eta,\eta\rangle =2$ where $\langle,\rangle$ is the obvious 
inner product for forms. 
The volume form is then given by $dV = \frac 12 \eta^2$.
In particular, it follows that $\eta$ is {\em selfdual}:
By $su(3)$ invariance it suffices to check this at the north pole
$x_a = \d_{a,8}$. There $f_{ab8}$ is manifestly selfdual, and so is
$\eta f(x)$ for any function $f(x)$.

Furthermore, we claim that the 2-forms of the form
\beq
\a_2 = f_{abc} dx_a dx_b A_c(x) \quad \mbox{with} 
  \quad \langle \a_2,\eta\rangle =0
\eeq
span the space of anti-selfdual 2-forms.
To see this, note again that the space of such 2-forms
is invariant under $SU(3)$. It therefore 
suffices again to consider  the north pole, where due 
to the explicit form of $f_{a b c}$ the space is easily seen to be 
3-dimensional and anti-selfdual.
This will be useful in the context of instantons.

\section{Fuzzy $\C P^2_N$.}
\label{sec:fuzzyCP2}

We start by recalling the definition of fuzzy $\C P^2$\cite{stroh,bala},
in order to fix our conventions. In general, 
(co)adjoint orbits \eq{conj-classes} on $G$ can be quantized 
(see \cite{sniat}, and \cite{qfuzzybranes} for an alternative
approach for matrix Lie groups $G$) in terms 
of a simple matrix algebra
$Hom_\C(V_N)$, where $V_N$ are suitable representations of $G$. 
The appropriate representations 
$V_N$ can be identified 
by matching the spaces of harmonics, i.e. the decomposition into
irreps under the symmetry group $G$ \cite{qfuzzybranes}. 
For $\C P^2$,  the correct harmonics are reproduced
for\footnote{alternatively one could use 
$V_{N\L_1} = V_{(N,0)}$, which gives an equivalent algebra
but a different embedding of $\C P^2 \subset \R^8$. Our choice matches
the classical geometry in Section \ref{sec:class-geom}.}
$V_N =  V_{N\L_2} = V_{(0,N)}$, which is the irrep of $su(3)$ with 
highest weight $N \L_2$.
Then the space of ``functions'' on fuzzy 
$\C P^2$ decomposes as 
\beq
\C P^2_N := Mat(V_{(0,N)}) =  
V_{(0,N)} \tens V_{(0,N)}^* = \oplus_{n=0}^{N} V_{(n,n)}.
\label{CPN-decomp}
\eeq 
under the (adjoint) action of $SU(3)$.
This  matches the decomposition \eq{CP2-harmonics} of functions on 
$\C P^2$  up to the cutoff.
To identify the fuzzy coordinate functions, we denote with
\beq
\xi_a = \pi_{V_N}(T_a)
\eeq
the representation of the generators $T_a$ \eq{T-generators} 
of the Lie algebra $su(3)$ acting on $V_{N\L_2}$.
Now consider the $3D_N \times 3D_N$ matrix 
($D_N$ is the dimension \eq{DN} of $V_{N\L_2}$)
\beq
X = \sum_a \xi_a \tau^a
\label{C-intro}
\eeq
where $\tau_a$ are the conjugated Gell-Mann matrices. 
As shown in Appendix B \eq{chareq-fuzzy}, $X$ satisfies the 
characteristic equation 
\beq
(X-\frac{2N}3 )(X+\frac N3 +1) =0.
\label{chareq-1}
\eeq
On the other hand, \eq{tau-algebra} implies 
\beqa
X^2 &=& \xi_a \xi_b \;
  (\frac 23 \d^{ab} + \frac 12 (i {f^{ab}}_c + {d^{ab}}_c) \tau^c).
\eeqa
Together with \eq{chareq-1} and the fact that $\xi_a$ are 
representations of $su(3)$, this gives the coordinate form of the
algebra of functions on $\C P^2_N$,
\beqa
i f^{ab}_c \xi_a  \xi_b &=& -3 \; \xi_c, 
   \qquad [\xi_a, \xi_b] = \frac i2 f^{ab}_c \xi^c\; \label{defxi1}\\
g^{ab} \xi_a \xi_b    &=& \frac 13 N^2 +N, \label{defxi2} \\
d^{ab}_c \xi_a  \xi_b &=& (\frac{2N}3 +1) \; \xi_c.   \label{defxi3}
\eeqa
E.g. for $N=1$, we simply have $\xi_a = \frac 12 \tau_a$. 
One then defines 
the rescaled variables $x_i= (x_1, ... x_8)$ of $\C P^{2}_N$  as
\beq
x_a = \L_N \; \xi_a, \qquad 
\L_N = R \frac{1}{\sqrt{\frac 13 N^2+N}}
\eeq
which satisfy the relations \cite{bala}
\beqa
i f^{ab}_c x_a x_b &=& -3 \L_N\; x_c = -3\frac{R}{\sqrt{\frac 13 N^2 +N}}\;x_c,
      \label{def1}\\
g^{ab} x_a x_b    &=& R^2, \label{def2} \\
d^{ab}_c x_a  x_b &=& R\;\frac{2N/3+1}{\sqrt{\frac 13 N^2 +N}}\; x_c. 
\label{def3}
\eeqa
Here $R$ is an arbitrary radius, which will usually be 1 in this paper.
Furthermore,
\beq
D_N := dim(V_{(0,N)}) = (N+1)(N+2)/2
\label{DN}
\eeq
from Weyl's  dimension formula, hence
the algebra of functions on fuzzy $\C P^2_N$ 
is simply $Mat(D_N, \C)$. 

The decomposition \eq{CPN-decomp} of functions into harmonics defines 
an embedding of the spaces 
\beq
\C P^2_N \hookrightarrow \C P^2_{N+1} \hookrightarrow ...  
\hookrightarrow \cC^{\infty}(\C P^2)
\label{embeddings-eq}
\eeq
(not as algebras), by
matching the harmonics of $su(3)$. Moreover, there is a 
corresponding equivariant quantization
map  $T^N: \cC^{\infty}(\C P^2) \to \C P^2_N$ \cite{stroh} obtained by
cutting off the higher modes, 
which makes precise 
how the algebras approach the classical one as $N \to \infty$. 
This allows
to measure the ``difference'' between fuzzy and 
classical functions using the 
operator norm resp. supremum norm, and statements like 
$f \in \C P^2_N \to f \in \cC^{\infty}(\C P^2)$ as $N \to \infty$ 
are understood in this sense throughout this paper.

\paragraph{Additional structure.}
We can easily identify a ``north pole'' in the fuzzy case\footnote{for
$V_{(N,0)}$ there would be a south pole and a north sphere}.
Indeed $\xi_8$ and $\xi_3$ can be simultaneously diagonalized, 
and the highest weight state $|N\L_2\rangle$ of $V_{(0,N)}$ has eigenvalues
$\xi_8 |N\L_2\rangle = \frac N{\sqrt{3}}\; |N\L_2\rangle$ and
$\xi_3 |N\L_2\rangle = 0$. This is the unique vector in $V_{(0,N)}$
with this maximal eigenvalue of $\xi_8$.
It is therefore natural to identify the delta-function on the
north pole with the projector
$|N\L_2\rangle\langle N\L_2 |$, 
i.e. to consider $\langle N\L_2 | f(x) |N\L_2\rangle$
as value of $f(x) \in \C P^2_N$ ``at the north pole''. For large $N$,
the eigenvalue of $x_8$ approaches $R$ as it should.

We can similarly find a fuzzy ``south sphere'' corresponding to the
sphere in \eq{southsphere}. Indeed, consider the subspace of $V_{(0,N)}$
with minimal eigenvalue $-\frac N{2\sqrt{3}}$ of $\xi_8$.
It is isomorphic to a $N+1$-dimensional irrep of the $su(2)$
subalgebra generated by $\xi_{1,2,3}$, hence it is a fuzzy sphere
with $\xi_1^2 + \xi_2^2 + \xi_3^2 = \frac{N(N+2)}4$. Therefore
$x_1^2 + x_2^2 + x_3^2 = \frac 34$ for large $N$ as in \eq{southsphere}.

The ``angular momentum'' operators (generators of $SU(3)$) become now inner,
\beq
L_a f(x) = [\xi_a,f],
\eeq
because then $L_a x_b = [\xi_a,x_b] = \frac i2 {f_{ab}}^c x_c$,
as classically. 
The integral on $\C P^2_N$ is defined by the suitably normalized trace,
\beq
\int f(x) % = \frac 2{(N+2)(N+3)}\; Tr(f) 
= \frac 1D_N Tr(f)
\eeq
and is invariant under $SU(3)$.

As a final remark,
let us reconsider the characteristic
equation \eq{chareq-1}
with eigenvalues $x = -\frac N3 -1$ resp. $x = \frac{2N}3$.
Their multiplicities $n_\pm$ can be obtained from 
$Tr(X) =0 = n_- (-\frac N3 -1) + n_+ (\frac{2N}3)$, which implies 
\beq
n_- = N(N+2), \qquad n_+ = \frac 12 (N+2)(N+3).
\eeq
This motivates to introduce the projector
\beq
P = \frac{X+\frac N3 +2}{N+1}
\eeq
which satisfies
\beq
P^2 = P, \qquad Tr(P) = \frac 12 (N+2)(N+3) \approx D_N.
\eeq
This clearly
generalizes \eq{projector-class} to the noncommutative case.

\subsection{Scaling limit: quantum plane}
\label{sec:plane}

Consider the coordinates 
\beq
z_a = \frac{\xi_a}{\sqrt{N}}
\label{z-def}
\eeq
for large $N$.
The commutation relations are 
\beqa
[z_a, z_b] &=& \frac {i}{2{\sqrt{N}}} f^{ab}_c z^c\; \label{defz1}\\
g^{ab} z_a z_b    &=& \frac N3  +1, \label{defz2} \\
z_c &=& \frac 3{2\sqrt{N}}\; \frac 1{1 + \frac 3{2N}}\; d^{ab}_c z_a  z_b.
     \label{defz3}
\eeqa
We are interested in the neighborhood of the ``north pole'', where
$z_8 \approx \sqrt{N/3}$ and $z_\mt = o(1)$.
Then \eq{constraint-explicit}
can be used to solve for $z_{1,2,3}$,
which implies that $z_{1,2,3} = o(\frac{z_\mt^2}{\sqrt{N}})$. 
Then \eq{defz1} implies
\beq
[z_4,z_5] = -\frac i2 + o(\frac{z_\mt^2}N),\qquad
[z_6,z_7] = -\frac i2 + o(\frac{z_\mt^2}N).
\eeq
in the large $N$ limit, we obtain 
$\R^4_\theta$ with 
\beq
\theta_{ij} = \left(\begin{array}{cccc} 0&1&0&0\\-1&0&0&0\\
                                   0&0&0&1\\0&0&-1&0 \end{array}\right)
\eeq

\section{Differential calculus}
\label{sec:diff-calc}

A differential calculus on $\C P^2$ was introduced in \cite{wataCPN},
which we recall and somewhat extend here.
We introduce a basis of one-forms $\theta_a$, $a=1,2,...,8$ \`a la Madore
\cite{Madore1991}, which transform in the 
adjoint of $su(3)$ and commute with the algebra of functions:
\beq
[\theta_a,f] =0, \qquad \theta_a \theta_b = - \theta_b \theta_a.
\eeq
This defines a space of exterior forms on fuzzy $\C P^2_N$, which we 
denote by $\Omega^*_N:= \Omega^*(\C P^2_N)$. The 
gradation given by the number of anticommuting generators $\theta_a$.
The highest non-vanishing form 
is the $8$-form 
corresponding to the volume form of $SU(3)$.

One can also define an exterior derivative 
$d: \Omega^k_N \rightarrow  \Omega^{k+1}_N$ such that $d^2 =0$ and imposing 
the graded Leibniz rule. Its action on the algebra elements 
$f \in \Omega^0_N$ is given by the commutator with a special one-form:
Consider the invariant one-form
\beq
\Theta = \xi_a \theta_a.
\eeq
The exterior derivative of a function $f \in \C P^2_N$ is then given by
\beq
d f := [\Theta,f] = [\xi_a,f] \theta_a.
\eeq
In particular, we have
\beq
d\xi_b =  [\xi_a,\xi_b] \theta_a = \frac i2 f_{abc} \theta_a \xi_c.
\label{dx}
\eeq
Note that for large $N$, this 
can be expressed in terms of the map $J$ as 
\beq
d x_a =  \frac{i\sqrt{3}}2 J(\theta)_a.
\label{dxi}
\eeq
This means that 
only tangential forms survive in the commutative limit, 
while the transversal forms becomes zero.
Therefore this calculus reduces to the usual calculus on $\C P^2$ 
for large $N$.
In particular, we  define
\beq
\langle \theta_a,\theta_b\rangle = -\frac 43 \d_{ab}
\label{inner-theta}
\eeq
which for large $N$ implies $\langle dx_a,dx_b\rangle = \d_{ab}$.
Furthermore, we define
\beq
 \eta = -\frac{\sqrt{3}}{8} f_{abc} x_a \theta_b \theta_c
\label{symplectic-NC}
\eeq
which satisfies $\langle\eta,\eta\rangle =2$ for any $N$.
For $N \to \infty$, it reduces to the symplectic form \eq{eta-class}
\beq
\eta \to -\frac{\sqrt{3}}{8} f_{abc} x_a J(\theta_b) J(\theta_c) = 
 \frac 1{2\sqrt{3}} f_{abc} x_a dx_b dx_c.
\eeq
The definition of $d$ on higher forms is straightforward, once we have
$d: \Omega^1_{N} \rightarrow \Omega^2_{N}$ such that $d^2(f) =0$.
 Notice \cite{qfuzzybranes} 
first that  there is a natural bimodule-map 
from one-forms to 2-forms, given by
\beq
\star_1(\theta_a) := \frac i4 f_{abc} \theta_b \theta_c.
\eeq
Using this, we define
\beqa
d: \; \Omega^1_{N} &\rightarrow& \Omega^2_{N}, \nn\\
                   \a~~ &\mapsto& d\alpha= [\Theta,\a]_+ - \star_1(\a)
\label{d_1}
\eeqa
where $\alpha\in\Omega^1_N$.
One can then verify $d^2=0$ in general. To see this, note that
$$
d df = [\Theta,df]_+ -\star_1(df)=0
$$ 
using the following relation:
\beq
\star_1(\Theta) = \Theta^2.
\label{starTheta}
\eeq
This follows from
\beq
\Theta^2 = \Theta \Theta =\frac 12 \theta_a \theta_b[\xi_a, \xi_b] 
 = i\frac 14 f_{abc} \theta_a \theta_b\xi_c 
  =-\frac{2i}{\sqrt{3}\Lambda_N} \eta.
\label{theta-2}
\eeq
One can also show that
\beq
d \Theta = \Theta^2, %= -\frac 2{n+1} d\xi_b d\xi_b,
\eeq
which implies
\beq
d\eta = 0.
\eeq

\paragraph{Field strength}

For an arbitrary one-form 
\beq
C = C_a \theta_a
\label{C-form}
\eeq
we define the field strength by
\beq
F:= C C - \star_1 C = -\frac i2 F_{ab} \theta_a \theta_b.
\label{F-formdef}
\eeq
Upon writing 
\beq
C = \Theta + A
\eeq
this becomes
\beq
F = dA + AA.
\label{F-A}
\eeq
Since the differential calculus reduces to the classical one for large
$N$, this  reduces indeed to the correct expression for the field strength
provided the ``gauge fields'' $A$ are purely tangential. 
How to implement this requirement will be discussed in Section 
\ref{sec:nc-constraints}.
The inner product of forms \eq{inner-theta} extends
as usual:
\beq
\langle \theta_a, \theta_b\rangle := -\frac 43 \d_{ab},
\qquad \langle \theta_a\theta_b, \theta_c\theta_d\rangle :=
\frac{16}9 (\d_{ac}\d_{bd} - \d_{ad} \d_{bc}),
\label{2forms-inner}
\eeq
and one can define an analog of the Yang-Mills action by
\beq
S_{YM} = Tr \langle F,F\rangle.
\label{YM-forms}
\eeq

One could now proceed and define the integrals of forms.
Here one meets an obstacle: there is no constant invariant 4-form
which could define the volume-form on fuzzy $\C P^2$. The most natural
candidate $d^4V = \eta^2$ does not commute with functions, and is
therefore not useful for gauge theory. Correspondingly, there seems
to be no natural notion of a (gauge-invariant) Hodge star. This
somewhat obscures the meaning of instantons in the fuzzy case, lacking
the concept of a self-dual field strength. Therefore these notions 
only make sense in the commutative of large $N$ limit.
The construction of topological invariants 
suffers from the same problem. This will mean that we will
calculate e.g. Chern numbers only in the commutative limit. This
problem
is well-known for fuzzy spaces, and perhaps not too surprising since
the notion of topology for spaces defined in terms of finite algebras
can only arise asymptotically.
We discuss in the following paragraph how such ``asymptotic'' 
Chern numbers can be computed.

\subsection{Asymptotic Chern numbers}
\label{subsec:chern}

The de Rham cohomology of classical $\C P^2$ is given by 
$H^2(\C P^2) = \R \eta$
and $H^4(\C P^2) = \R \eta^2$. The integer cohomology $H^{2*}(\C P^2; \Z)$
is generated by
\beq
\omega =  \frac{\eta}{3\pi},
\label{omega}
\eeq
i.e. 
\beq
\int_{S^2_S} \omega = 1 = \int_{\C P^2}
\omega \wedge \omega
\label{cycle-normalization}
\eeq
where the 2- and 4-cycles are
represented by the south sphere $S^2_S$ discussed in
Section \ref{sec:class-geom}, and $\C P^2$ itself. 
We therefore expect 2 interesting Chern 
classes $c_1 = \frac i{2\pi}\; tr F$ and
$c_2 = -\frac 1{8\pi^2} (tr F \wedge tr F - tr(F\wedge F))$.
The first Chern number is given by the integral over the south sphere
\beq
c_1 = \frac i{2\pi}\; \int_{S^2_S} tr F
 = \frac i{2\pi}\; \int_{S^2_S} \frac 12 tr \langle F,\eta \rangle
\label{c-1}
\eeq
since $\langle \eta,\eta\rangle = 2$.
Now 
\beq
\langle F,\eta \rangle 
 = \frac{\sqrt{3}i}{16} F_{ab}  f_{def} x_d
   \langle \theta_a \theta_b,  \theta_e \theta_f  \rangle
= \frac{2i}{3\sqrt{3}}  F_{ab}  f_{abc} x_c
= \frac{4i}{3\sqrt{3}}  F_{c} x_c
 \approx \frac{4 i}{3 N}  F_c C_c
\label{F-eta}
\eeq
where $F_a$ is defined as in 
\eq{Fa-def}. Here ``$\approx$'' means equal up to $o(1/N)$.
This will be evaluated below for the configurations of
interest.
Another way to compute $c_1$ 
which is especially interesting for the noncommutative case
is to note that $\eta \wedge c_1 \in H^4(\C P^2) \cong \R$, hence  
\beqa
c_1 &=& \frac 1{3\pi} \int_{\C P^2} \eta \wedge c_1 
     = \frac i{6\pi^2} \int_{\C P^2} \eta \wedge tr F  \nn\\
 &=& \frac i{6\pi^2} \int_{\C P^2} \star \eta \wedge tr F
 =  \frac i{6\pi^2} \int_{\C P^2} tr  \langle \eta,F \rangle d^4 V \nn\\
 &\approx& -\frac 2{9\pi^2 N} \int_{\C P^2} tr  F_c C_c d^4 V 
 = -\frac 1{N D_N} Tr  F_c C_c  
\label{c1-4dim}
\eeqa
for large $N$. 
The integral is now over the entire $\C P^2$, and
$d^4 V = \frac 12 \eta^2$ is the volume form on $\C P^2$.
Here we use the metric 
$d^2s = \sum_{a=1}^8 dx_a dx_a$; this
leads to the volume
$Vol(\C P^2) = \int_{\C P^2} d^4 V = \frac 92 \pi^2$.
We note that the last expression 
in \eq{c1-4dim} has a topological meaning, which will be important later
as a possible term in the action of a gauge theory.

\section{Multi-Matrix Models and Yang-Mills action}
\label{sec:YM}

\subsection{Degrees of freedom and field strength}

Our basic degrees of freedom are 8 hermitian matrices $C_a \in
Mat(D_N,\C)$ transforming in the adjoint of $su(3)$, 
which are naturally arranged as a single $3 D_N \times 3 D_N$ matrix
\beq
C =  C_a \tau^a + C_0 \one
\eeq
where $C_0 =0$ in much of the following.
The size $D_N$ \eq{DN} of these matrices will be relaxed later.
We want to find a multi-matrix model in terms these $C_a$,
which for large $N$ reduces to (euclidean) Yang-Mills 
gauge theory on $\C P^2$. The idea is to interpret the $C_a$ as
suitably rescaled ``covariant coordinates'' \cite{wess} 
on fuzzy $\C P^2_N$, with the
gauge transformation
\beq
C_a \to U^{-1} C_a U
\eeq
for unitary matrices $U$ (of the same size). The $C_a$ can also be
interpreted as components of a one-form $C = C_a \theta_a$ as in \eq{C-form}.
Following the approach of \cite{matrixsphere}, we look for an action which 
has the ``vacuum'' solution 
\beq
C_a = \xi_a
\eeq
corresponding to $\C P^2_N$, and forces $C_a$ to be ``approximately''
the corresponding representation $V_{N \L_2}$ of $su(3)$.
Then the fluctuations 
\beq
C_a = \xi_a + A_a
\eeq
are small and describe the gauge fields. By inspection,
these gauge fields $A_a$ transform as 
\beq
\d A_a = i[\xi_a + A_a, \Lambda] 
=  i L_a \Lambda + i [A_a,\Lambda]
\eeq
for $U =e^{i\L}$,
which is the appropriate formula for a gauge transformation.
Since the $C_a$ resp. $\xi_a$ 
correspond to ``global'' coordinates in the embedding space 
$\R^8$, we can hope that nontrivial solutions such as instantons can
also be described in this way.

A suitable definition for the field strength is then given by
\eq{F-formdef},
\beq  \fbox{$
F_{ab} = i[C_a,C_b] + \frac 12\; f_{abc} C_c 
      = i (L_a  A_{b} - L_b A_a +  [A_a,A_b]) + \frac 12\; f_{abc}
      A_c.
$}
\label{fieldstrength}
\eeq
We will also need
\beq  \fbox{$ \begin{array}{ll}
\vspace{5pt}
F_a & = \;\; i f_{abc} C_b C_c + 3 C_a = \frac 12 f_{abc} F_{ab}, \\
 D_a &= \;\; d^{ab}_c C_a  C_b - (\frac{2N}3 + 1)\; C_c.  \end{array} $}
\label{Fa-def}
\eeq
Under gauge transformations, the field strength transforms as 
\beq
F_{ab} \to U^{-1} F_{ab} U.
\eeq
As discussed in Section \ref{sec:diff-calc}, 
this can also be interpreted as 2-form
\beq
F = dA +AA   = \frac{2i}3 F_{ab}  J(dx_a)  J(dx_b) 
\label{F-form}
\eeq
using  \eq{dxi}, \eq{F-formdef},
if one considers the fields $C_a$ as one-forms
$C = C_a \theta_a = \Theta + A$. 
Furthermore, we will show that
$F_{ab}$ is tangential if $C_a$ satisfies 
(approximately) the constraints of $\C P^2$. 
Assuming that $A_a$
tend to well-defined functions on $\C P^2$ in the large $N$ limit and
using \eq{dxi}, this implies that  
$F_{ab}$ are the components of the usual 
field strength 2-form in the commutative
(large $N$) limit, transformed by $J$ acting on the indices.
This justifies the above definition of $F_{ab}$, and it is a matter 
of taste whether one works with the components or with forms.
Note that our $C$ and $A$ are dimensionless, and the usual 
dimensions are recovered upon introducing a radius $R$ and suitably
rescaling the quantities $A_a \to R A_a'$ etc.

It is instructive to consider $F_a$ explicitly: at the north pole,
its components are
\beqa
F_1 &=& {i}((L_4 A_7 - L_7 A_4) - (L_5 A_6 - L_6 A_5))  \nn\\
F_2 &=& {i}((L_4 A_6 - L_6 A_4) + (L_5 A_7 - L_7 A_5)) \nn\\
F_3 &=& {i}((L_4 A_5 - L_5 A_4) - (L_6 A_7 - L_7 A_6)) \nn\\
F_8 &=& -{i}\sqrt{3} ((L_4 A_5 - L_5 A_4) + (L_6 A_7 - L_7 A_6)).
\eeqa
We see explicitly that $F_{1,2,3}$ are antiselfdual, while $F_8$
is selfdual.

\subsection{Constraints}
\label{sec:nc-constraints}

In order to describe fuzzy $\C P^2$,
the fields $C_a$ should satisfy at least approximately the
constraints \eq{defxi2}, \eq{defxi3} of $\C P^2_N$,
\beqa
D_a &=& 0,  \label{defC3} \\
g_{ab} C_a C_b &=& \frac 13 N^2 +N \label{defC2} 
\eeqa
which are gauge invariant.
These constraints ensure that $C_a$ can 
be interpreted as describing a (``dynamical'' or fluctuating) $\C P^2_N$. 
However, notice that they are not independent 
at least in the commutative limit
as was shown in Section \ref{sec:tensor-maps}: \eq{def3c} implies 
\eq{def1c}.
To understand these constraints in the noncommutative case, we introduce some 
analogs of the maps in Section \ref{sec:tensor-maps}.

\subsubsection{Linear tensor maps}
\label{sec:linmap-NC}

Given 8 matrices $C_a$ as above, we can define
the following  maps of matrices $X_a$ with index in the adjoint
of $su(3)$:
\beqa
\cJ(X)_a &=& \frac 1{2N}\;  f_{abc} \{C_b, X_c\} ,   \nn\\
\cD^{lin}(X)_a &=& \frac 1{2N}\;  d_{abc} \{C_b, X_c\} - (\frac 13
+\frac 1{2N}) X_a  \nn\\
\cP(X)_a &=& \frac 1{\frac 43 N^2 +4N} \{C_a, \{C_b, X_b\}\}.
\eeqa
They map hermitian matrices into hermitian matrices, and depend
on the particular ``background'' field $C$.
Note that $\cJ$ is anti-selfadjoint in the following sense
\beq
Tr(\cJ(X)_a Y_a)   = - Tr(X_a \cJ(Y_a))
\eeq
while $\cD^{lin}$ is selfadjoint.
Writing $C_a = \xi_a + A_a$, these maps reduce to the
corresponding maps in Section \ref{sec:tensor-maps} in the commutative limit
\beq
\cD^{lin}(X)_a \to D^{lin}(X)_a,  \qquad
\cJ(X)_a  \to J(X)_a  ,  \qquad \cP(X)_a \to P(X)_a
\label{cJ_approx}
\eeq
as $N \to \infty$ independent of $C_a$, 
provided $A_a$ and $X_a$ are ``smooth''. With
smooth we mean that both $X_a$ and
its derivatives $[\xi_b,X_a]$  become smooth functions as $N \to
\infty$, but mild singularities may be allowed.
Therefore tangential tensors could be described by solutions of 
$\cD^{lin}(X)_a =0$. However this seems to be too restrictive in the
noncommutative case, and
we will only require
\beq
\cD^{lin}(X)_a \to 0 \quad\mbox{ as} \quad N \to \infty.
\label{linear-constraint}
\eeq
Recall that this is to be understood in the sense of \eq{embeddings-eq}.
Alternatively, 
they can also be described by the image of $\cJ(X)$, which is
equivalent for large $N$.

Using \eq{cJ_approx}, the maps $\cD^{lin}$ and $\cP$ 
can be used to project 
any tensor field to its tangential components for large $N$ 
as in \eq{tang-proj-approx-class}, by 
\beq
X \to X_{tang} = X - \frac 43 \cP(X) + \cD^{lin}(X).
\label{tang-proj-approx}
\eeq
The resulting field $X_{tang}$ satisfies \eq{linear-constraint}, which
is enough for our purpose. Therefore imposing
\eq{linear-constraint} allows 4 tangential degrees of freedom in the 
large $N$ limit.

\subsubsection{Nonlinear tensor map}
 
One can also consider the noncommutative versions of the map in Section 
\ref{subsubsec:nonlinmaps}:
\beqa
\cD^{nl}(C_a) &=&  \frac 1N\; D_a 
   = \frac 1N\;(d_{abc} C_b C_c  -(\frac{2N}3+1) C_a),
\eeqa
which maps hermitian matrices into hermitian matrices.
For infinitesimal variations, we have 
$\d \cD^{nl}(C) =  \cD^{lin}(\d C)$.
It would be tempting to impose the constraint $\cD^{nl}(C_a) =0$.
However, it is not clear whether this equation has 
enough solutions in the noncommutative case to describe 4 tangential vector
fields. As this is probably not be the case, we will not pursue this strict
constraint any further, and we only require
\beq
\cD^{nl}(C)_a \to 0 \quad\mbox{ as} \quad N \to \infty.
\label{nonlinear-constraint}
\eeq
This guarantees the correct commutative limit, and it is equivalent to
\eq{linear-constraint} as long as $A_a$ is ``smooth''
since then $\cD^{nl}(C)_a = \cD^{lin}(A)_a + o(\frac 1N)$. Therefore
\eq{nonlinear-constraint} admits 4 tangential degrees of freedom in the 
large $N$ limit via \eq{tang-proj-approx}. Moreover, \eq{nonlinear-constraint}
holds for finite-action configurations of the 
Yang-Mills action defined below.

Nevertheless, there exists a slightly modified constraint 
which can be imposed exactly, which will be discussed in Section 
\ref{sec:constraints}.

\subsubsection{Constraints and field strength}
\label{sec:constfield}

We can now verify that $F_{ab}$ is (approximately) tangential
in the sense that $\cD^{lin}(F) \to 0$ for each index  as $N \to \infty$,
provided both $\cD^{nl}(C)_a \to 0$ and 
 its derivatives $[C_a,\cD^{nl}(C)_b] \to 0$ for $N \to \infty$. 

Generalizing the above definition to higher tensors, consider 
\beqa
\cD^{lin}_{1}(F_{ad}) &=& \frac 1{2N} d_{axy} (C_x F_{yd} + F_{yd}  C_x)
     - (\frac 13 +\frac 1{2N}) F_{ad} \nn\\
 &=&  \frac 1{2N} d_{axy} \Big(i(-C_x C_d C_y + C_y C_d C_x
     +  C_x C_y C_d - C_d  C_y C_x) \nn\\
 &&  + \frac 12( f_{ydz} C_x C_z + f_{ydz} C_z C_x)\Big) 
    - (\frac 13 +\frac 1{2N}) F_{ad}   \nn\\
&=&  \frac 1{2N} \Big([i(d_{axy} C_x C_y),C_d]
   + \frac 12( f_{dce} d_{eab} + f_{dbe} d_{eac} ) C_b C_c\Big)
      - (\frac 13 +\frac 1{2N}) F_{ad}   \nn\\
&=& \frac 1{2N} \Big(i[(d_{axy} C_x C_y),C_d]
   + \frac 12 f_{ade} (d_{ebc} C_b C_c)\Big)  
    - (\frac 13 +\frac 1{2N}) F_{ad}  \nn\\
&=& \frac 1{2N} (i[D_a,C_d]
   + \frac 12 f_{ade} D_e )  = o(\cD^{nl}(C)) + o([C_a,\cD^{nl}(C)_b]) 
\label{F-tang}
\eeqa
using \eq{fd-identity}. The subscript $\cD^{lin}_{1}$ indicates that 
$\cD^{lin}$ is applied to the first index.
Hence $F_{ab}$ is indeed tangential with the above assumptions. 
Furthermore, note that the last term $f_{abc} A_c$
in the definition \eq{fieldstrength} of $F_{ab}$ 
does not contribute for tangential
$A_a$. Then the usual formula for $F_{ab}$
is reproduced e.g. at the north pole.

\subsection{The Yang-Mills action}
\label{sec:YM-action}

Assume that the  $C_a$ satisfy the constraints \eq{defC3} 
(and therefore also \eq{defC2}) of $\C P^2_N$
exactly or approximately, so that 
$F_{ab}$ is tangential as shown above.
Then the ``Yang-Mills'' action is defined as
\beq
S_{YM} = \frac 1{g }\;  \int  F_{ab} F_{ab}
  =  \frac 1{g D_N}\; Tr(-[C_a,C_b]^2 +2if_{abc}C_aC_bC_c +3C_a C_a).
\eeq
It reduces to the classical Yang-Mills action on $\C P^2$, because only the
tangential indices contribute as shown in \eq{F-tang}. $S_{YM}$ can also be
written in the form \eq{YM-forms}.
The corresponding 
equation of motion is $-2 [C_b,[C_a,C_b]] +3i f_{abc} C_bC_c +3
C_a = 0$, which is
\beq
2 [C_b,F_{ab}] - i F_a  = 0 
\label{eom-first}
\eeq
The last term may seem strange, but it does not contribute 
to the tangential fields since $F_{\mt} \to 0$ as $N \to \infty$ by 
\eq{F-tang}. 
Noting that $[C_b,.] = [\xi_b + A_b,.]$ corresponds to the covariant
derivative in the adjoint, 
this is equivalent to the usual equation $D^\mu F_{\nu \mu} =0$ for the
tangential fields.

We now have to impose the constraints \eq{defC3}, \eq{defC2}
either exactly or approximately, and
there are several possibilities how to proceed.
Imposing both of them exactly seems too restrictive; 
recall that they are not independent even classically.
One can hence either 
\begin{enumerate}
\item
consider all 8 fields $C_a$ as dynamical and add
\beq
S_{D} = \frac 1{g D_N}\; Tr \( \mu_1 (dCC-(\frac{2N}3 + 1)C)^2 +
\mu_2(C\cdot C - (\frac{N^2}3 +N))^2\)
\label{S_cons}
\eeq
to the action. This will impose the constraint dynamically for
suitable $\mu_1 > 0$ and $\mu_{2} \geq 0$, by giving the 4 transversal
fields a large mass $m \to \infty$. Or,
\item
impose the constraint $D = dCC-(\frac{2N}3 + 1)C =0$ 
exactly, or a slightly modified version.
\end{enumerate}
In the second approach, it is not clear
whether there are sufficiently many solutions of 
$D=0$ in the noncommutative case to admit 
4 tangential gauge fields. This concern
could be circumvented by modifying the constraint, which  
is discussed in Section \ref{sec:constraints}. 
However we have not been able to find instanton-like solutions in this
case (which may just be a technical problem). Therefore we
concentrate on the first approach here, where we do find  
topologically nontrivial
instanton solutions. 
It offers the additional possibility 
to give physical meaning to the non-tangential degrees of freedom.

Therefore our action is
\beq \fbox{$
S = S_{YM} + S_D$.}
\label{S-gauge-final}
\eeq
We will show  that this reproduces the
classical Yang-Mills action on $\C P^2$ in the large $N$ limit for 
\beq\fbox{$
\mu_{1} = o(\frac 1N), \quad \mbox{and}\quad \mu_2 \leq o(\frac 1{N^3}).$}
\label{mu-range}
\eeq
Here $o(\frac 1N)$ stands for a function which scales like 
$\frac 1N$.
We proceed to find the ``vacuum'', i.e. the minimum
of the action. Assume first that the size of the matrices is $D_N$;
this will be somewhat relaxed below. Then 
the absolute minima of the action are given by solutions of 
$F_{ab} = 0$ and $D_a = 0$, 
which means that $C_a$ is a representation of
$su(3)$ with $D_a = 0$. The latter implies (e.g. using \eq{char-X}) 
that the only allowed irreps are 
$V_{N \L_2}$ or the trivial representation, and the trivial one is
suppressed by $\mu_2$.
To determine the appropriate range for $\mu_{1,2}$, note that 
$\mu_1 = o(1/N)$ is sufficient by \eq{D-term} to decouple the 4 transversal
degrees of freedom. It implies that $\cD^{nl}(C) \to 0$ for $N \to
\infty$ (for finite action) 
as required in Section \ref{sec:linmap-NC}, hence it allows 4
tangential gauge fields. On the other hand,
the instanton solutions found 
in Section \ref{sec:top-nontriv} have $D_a = o(1)$ and also
$C\cdot C - (\frac{N^2}3 +N) = o(1)$, therefore we need
$\mu_{1,2} \to 0$ to make sure $S_D$ does not contribute to the
Yang-Mills action in the commutative limit $N \to \infty$. Finally, we need
$\mu_2 \leq o(1/N^3)$, otherwise the trivial blocks in 
\eq{monopole-matrix} would
be suppressed. All this leads to \eq{mu-range}. One might allow
$\mu_2 =0$ and  argue that the nontrivial vacuum is chosen 
due to the larger phase-space.  

Therefore the ``vacuum'' solution is 
\be 
(C_{vac})_a = \xi_a
\label{C-vac}
\ee 
in a suitable basis, and other saddle points 
have a larger action.
These arguments go through if we allow the
size of the matrices $C_a$ to be somewhat bigger
that $D_N$, say
\beq
C_a \in Mat(D_N + N, \C)
\eeq 
(anything much smaller that $2 D_N$ will do), which is needed to
accomodate all the nontrivial solutions found below. 
Any configuration with finite action is therefore close to 
\eq{C-vac}, and can hence be written as 
\beq   
 C_a  = \xi_a \;  + A_a
 \label{fluc-C}
\eeq 
in a suitable basis, with ``small'' $A_a$.
This justifies the assumptions made in the beginning of  
Section \ref{sec:YM-action}.

\paragraph{Coupling to matter fields.}

To further clarify the physical meaning of this matrix-model
approach to gauge theory,
assume that we have an additional complex scalar field
$\phi$. Without gauge coupling, a natural action would be 
$\int ([\xi_a,\phi])^\dagger [\xi_a,\phi] = -\int \phi^\dagger \Delta \phi$.
If we assume that $\phi$ is charged and
transforms under gauge transformations as
\beq
\phi \to U \phi,
\eeq
then a natural gauge-invariant action would be 
\beq
S[\phi] = \int (C_a \phi - \phi \xi_a)^\dagger (C_a \phi - \phi \xi_a). 
\label{phi-action}
\eeq
This reduces to $\int (D_a \phi)^\dagger D_a \phi$
where $D_a = [\xi_a,.] + A_a$. The combined action $S_{YM} +S[\phi]$
is again a 
matrix model, and we should expect solutions where 
both $C_a$ and $\phi$ are block-matrices in $End(V)$. 
In particular, $\phi$ may effectively be a rectangular
matrix $Hom(V,V')$  (the rest being filled by zeros, 
say). This is exactly what happens for 
the monopole and instanton solutions constructed below,
where $C_a \in End(V') \subset End(V)$ couples naturally to 
$\phi \in Hom(V,V')$. The latter should therefore be considered as a
section in an associated (nontrivial) bundle, which will be 
made very explicit in  Section \ref{sec:class-bundle}. This is in complete
agreement with the results of \cite{grosse-topology,wataCPN}, 
establishing an indirect 
connection with the approach using projective modules. 

The appropriate action for fermions is not obvious since $\C P^2$ 
has no spin but spin${}^c$ structure. Some proposals for a Dirac
operator have been given in the literature \cite{stroh,bala}, and
we postpone the formulation in our approach to future work.

\paragraph{Rewriting the action and equations of motion.}

One important observation is that using some identities of the
$su(3)$ tensors (see \eq{YM-rewritten}, Appendix C), one can rewrite the 
YM term as
\beq 
S_{YM}=\frac 1{g D_N} Tr\(2(C\cdot C) (C\cdot C) -\frac 32(dCC)(dCC) 
    - \frac 12 (fCC) (fCC) +2if C C C +3 C \cdot C\)
\label{YM-rewritten1}
\eeq
where the obvious index structure has been omitted, and $C \cdot C = C_a C_a$.
Using this and
\beqa
Tr((\d (f C C)) X_a) &=& Tr(\d C_a(f_{abc} [C_b,X_c])),  \nn\\
Tr((\d (d C C)) X_a) &=& Tr(\d C_a(d_{abc} \{C_b,X_c\}))
\eeqa
the equations of motion for $S_{YM}$ can be written as
\beq
 4 \{C_a, (C\cdot C)\}  +6 C_a -3 d_{abc} \{C_b,(dCC)_c\}+ if_{abc}
[C_b,F_c] =0.
\label{eom-1}
\eeq
This is much easier to work with. 
$S_D$ leads to the additional term
\beq
 \mu_1\Big(2 d_{abc}\{C_b,(dCC)_c\}  -6(2N/3+1) d_{abc} C_b C_c  
  +2(2N/3+1)^2 C_a\Big) 
  + 2 \mu_2\{C_a, C\cdot C -(\frac{N^2}3 +N)\}
\label{eom-additional}
\eeq
in the lhs of  \eq{eom-1}.
A particularly interesting form \eq{S-VVp} is obtained for 
$\mu_1 =2, \mu_2 = -\frac 23$, which is however outside of the 
range \eq{mu-range} where the model is under control.

\subsection{Decoupling of auxiliary variables}
\label{sec:decoupling}

As discussed above, we impose the constraints of $\C P^2_N$ 
 by adding the term %\eq{S_cons}
\beq
S_{D}  = \frac 1{g D_N}\;  Tr\( \mu_1 D_a D_a 
   + \mu_2 (C\cdot C - (\frac{N^2}3 +N))^2\) 
\eeq
to the action. 
We will now show that this amounts to 
giving the 4 transversal fields a large mass $m \to \infty$ which 
therefore decouple, leaving 4 massless tangential gauge fields.
In fact one can put $\mu_2 =0$, since  \eq{defC2} is not an
independent constraint. Using
$C_a = \xi_a + A_a$
we get
\beqa
D_c &=&  {d^{ab}}_c \{\xi_a, A_b\} +  {d^{ab}}_c A_a  A_b 
          -  (\frac{2N}3+1) A_c  \nn\\
C_a C^a - (\frac{N^2}3 +N) &=& \xi_a A^a + A_a \xi^a +  A_a A^a.
\eeqa
In particular, assuming that $A_a$ and $[A,A]$ are ``smooth'' we have
\beq
\frac {D_c}{2N} = \frac 1{2N} {d^{ab}}_c \{\xi_a, A_b\} 
 - \frac 1{3}\;A_c + o(1/N) 
=  \cD^{lin}(A_c) \,\;+ o(1/N)
\eeq
If $A_a$ are (the
quantization \eq{embeddings-eq} of) regular classical tangential fields,  
we can assume that $\cD^{lin}(A_c) = o(1/N)$
using  results in \cite{stroh}.
Therefore $\frac {D_c}N \approx 0$ for such
tangential fields. On the other hand, $\cD^{lin}(A)$
reproduces the transversal fields 
with their respective eigenvalue $-1$ resp. $\frac 13$.
To see this explicitly, consider the ``north pole'', where 
$x_a \approx \d_{a,8}$. Then $A_\mt = A_{4,5,6,7}$ are tangential, and
$A_{1,2,3}$ and $A_8$ are ``transversal'' with
\beqa
\frac{D_{\mt}}{2N}    &=&  o(1/N), \nn\\
\frac{D_{1,2,3}}{2N}  &=&  -  A_{1,2,3}\; +o(1/N), \nn\\
\frac{D_8}{2N}        &=& \frac{1}{3} A_8 \; +o(1/N), \nn\\
\frac 1{2N}(C_a C^a - (\frac{N^2}3 +N)) &=& \frac 1{\sqrt{3}} A_8 + o(1/N).
\eeqa 
This shows how $D_a$ separates the tangential from the
transversal fields. Therefore the term $\mu_1 D_a D_a$ gives the 
transversal modes\footnote{the splitting 
into tangential and transversal modes is thereby defined in a global way} 
$A_\mr$  a mass term
of order $\mu_1 N^2$, while the tangential modes are affected by terms of 
order at most $\mu_1$. Similar conclusions can be drawn for $\mu_2$.
This means that if we choose 
\beq
\mu_{1} = 1/N,
\eeq
then 
\beq
\mu_1 D_a D_a = 4 N (A_1^2+A_2^2+A_3^1+\frac 19 A_8^2)
\label{D-term}
\eeq
(at the north pole), and similar for the other term. This implies that
 our action \eq{S-gauge-final} indeed 
approaches the classical Yang-Mills action for $N \to \infty$.

Note that for the special choice 
$\mu_1=2, \mu_2 = -\frac 23$ considered in Section \ref{sec:onetwo}, the mass
term for $A_8$ is undetermined. The meaning of the model is then unclear.

\paragraph{Additional terms in the action.}

Based on $su(3)$ invariance, 
one should also allow other terms such as 
\beq
\int a_1\; C\cdot C + a_2 \; fCCC + a_3\; dCCC
\eeq
etc in the action. However, recall that such $su(3)$ singlets 
may {\em not} be invariant under local
$so(4)$ rotations in the commutative limit. Consider first the terms 
$dCCC$ and $C\cdot C$. They are clearly related to the constraints 
\eq{defC3} and \eq{defC2}, hence they are essentially constants, 
and therefore harmless for small enough coefficients $a_i$. They are also
covariant in the usual sense,
which can be seen using the explicit form of the $d_{abc}$.
The term $fCCC$ is less obvious at first sight, since it is not 
covariant in the usual sense.
This may seem very dangerous: One might of course
exclude this term from the bare action, 
however quantum corrections are likely to 
restore it. 
Fortunately, it is again harmless:   
according to \eq{c1-4dim}, it essentially reduces to
the first Chern number (plus a constraint)
in the classical limit, which is topological and does not affect
the local physics as long
as $a_2 = o(1/N)$. Using \eq{F-eta}, we see that 
the explicit breaking of covariance is 
due to the symplectic form $\eta$ which lives on $\C P^2$.

We note in passing that in string theory or boundary CFT \cite{ARS}, 
fuzzy spaces arise as $D$-branes on group manifolds, and
a term of the form $ fCCC$ occurs in the effective action
(the ``Chern-Simons'' term). We now understand that 
this is a combination of the first Chern number 
and a Casimir-constraint. This interpretation
is expected to hold also for higher-dimensional branes.
However, the constraints \eq{defC3}, \eq{defC3} do not arise
in this context, hence the action in \cite{ARS} 
has a completely different
physical meaning.

\subsection{Alternative approach: constraints}
\label{sec:constraints}

As discussed above, imposing both
\eq{defC3}, \eq{defC2} in the fuzzy case is probably too restrictive. 
The first remedy is to drop \eq{defC2}, because it is redundant classically
and is not compatible with the monopole solutions below. 
Hence consider $D_a =0$ only. However, it appears that even this is 
too strong, at least we see no reason why these 8 equations 
should admit 4 independent
degrees of freedom; note that the kernel of
$\d \cD^{nl} = \cD^{lin}$ is probably smaller than classically. 
To find a way out, note first that a slightly modified constraint
\beq
D_a = d_{abc} C_b C_c -(\frac{2N}3+1)C_a = - F_a  
\label{constr-mod}
\eeq
would also be acceptable, because it implies  $\cD^{lin}(A) =  o(1/N)$ 
as long as $F$ is finite; this is all we needed above.
This suggests to consider the following slightly generalized constraint, 
which indeed admits 4 tangential degrees of freedom: 
introducing a further (auxiliary) scalar field $C_0$, consider
\beq
C = C_0 + C_a \tau^a,
\eeq
and impose the constraint
\beq
C^2 = \frac{(N+1)^2}4.
\label{C2-const}
\eeq
This means that $C$ has 
2 eigenvalues $\pm \frac{N+1}2$, and the multiplicities $n_\pm$ 
are determined by $Tr(C)$. 
In component form, this is equivalent to
\beq
\frac 23\; C_a C_a + C_0^2- \frac{(N+1)^2}4 =0 
 =(i {f_{abc}} + {d_{abc}}) C_a C_b + 4\{C_0, C_c\}.
\eeq
If $C_0 \approx const$
(which could be enforced by adding a term $[C_a,C_0]^2$ to the
action), then the first equation implies $C_0 \approx \frac{-N+3}6$,
and the second equation reduces to \eq{constr-mod}.
In particular, this is solved by
\beq
C = \frac{-N+3}6 + \xi_a \tau_a.
\eeq
Note that $C_0$ is essentially determined by the first equation, 
revealing its auxiliary nature.

It is easy to see that \eq{C2-const} admits 4 tangential degrees 
of freedom:
Indeed, the most general variation of some $C$ consistent with this
constraint has the form
\beq
\d C = [X,C] \approx i N \cJ(X)
\eeq
for an arbitrary matrix $X = X_a \tau_a$, cp. \eq{cJ_approx}. Since
$\cJ$ projects out the radial degrees of freedom, only the 
tangential fields survive and can be chosen freely.

A further advantage of this approach is the fact that the 
path integral is an integral over the compact space 
$U(3D)/(U(n_+)\times U(n_-))$. Also, it turns out that 
the monopole and instanton 
configurations below are  compatible with that
constraint\footnote{this motivates introducing $\tilde D$ in 
\eq{D1-def}, because it
satisfies a quadratic characteristic equation}.
All this is quite appealing.
However, these instantons turn out to be 
no longer solutions of the corresponding
equation of motion. 
The significance of this fact is not clear; it might 
be that there are other ``nearby'' solutions in this case, or it might
indicate some different global implications of this constraint. 
Furthermore, the addition of the field $C_0$ somewhat 
complicates the analysis.
We therefore concentrate on the approach with auxiliary variables.

\subsection{Nonabelian case: $U(n)$ Yang-Mills}

The generalization to a $U(n)$ gauge
theory is straightforward as in \cite{matrixsphere}, and is 
achieved by considering the same action for larger
matrices $C_a$. In order to  accomodate the monopole and instanton
solutions constructed below, we allow the size to be somewhat larger
that $n D_N$, say
\beq
C_a \in Mat(n (D_N + N), \C)
\eeq 
(anything much smaller that $(n+1) D_N$ will do).
Then consider the same matrix model \eq{S-gauge-final} as before.

The following analysis is parallel to that in Section \ref{sec:YM-action}.
First we should find the ground state. For $\mu_2=0$, 
the absolute minima of the action are given by solutions of $D_a = 0$
and $F_{ab} = 0$, which means that $C_a$ is a representation of
$su(3)$ with $D_a = 0$. The latter implies that the representation
decomposes into a direct sum of either 
$V_{N \L_2}$ or the trivial rep. The trivial \reps are suppressed by 
making $\mu_2$ slightly bigger than zero, and 
in a suitable basis $C$ takes the form 
\be %\label{ground-C-nonabel}
(C_{vac})_a = \xi_a\; {\bf 1}_{n\times n}
\label{C-vac-nonabel}
\ee 
which is a block matrix consisting of $n$ blocks of the fuzzy 
$\C P^2$ solutions.
The action is then zero, and clearly all other saddle points 
have a positive action. 
Any configuration with finite action is therefore close to 
\eq{C-vac-nonabel}, and can  be written as 
\beq   
 C_a  = \xi_a \;  + A_a
 \label{fluc-C-nonabel-2}
\eeq 
where  $A_a$ is ``small'' and carries an additional  $u(n)$ index,
\be 
A_a =  A_{a,\a}\; \la_{\a}.
\ee
Here $\la_\a$ denote the Gell-Mann matrices of $u(n)$,
and $\la_0 = \bf{1}$ is the $n\times n$ unit matrix. 
The rest of the analysis of the previous sections 
goes essentially through, in particular the transversal components
of $A_{a,\a}$ will be very massive and decouple due to $S_D$.
The field strength is again $F = dA + AA$ or \eq{fieldstrength}
in component notation,
which becomes the usual field strength of a $U(n)$ gauge theory
if the $u(n)$ indices are spelled out. In particular, all
non-tangential components are suppressed again due to 
\eq{F-tang}, and the equations of motion have the same form 
\eq{eom-first} or \eq{eom-1}, \eq{eom-additional}. 
All this will be understood in the following.

\subsection{One- and Two-Matrix Models for Fuzzy $\C P^{2}_N$}
\label{sec:onetwo}

It is tempting to consider also the analog of the single-matrix 
model studied in \cite{matrixsphere} for the fuzzy sphere.
Consider a hermitian $3 D_N \times 3 D_N$ matrix
\beq
C =  C_0 \one +  C_a \tau_a  =  \frac{-N+3}6\; \one +  C_a \tau_a.
\eeq
Then
\beq
C^2 =
   \(\frac 23 C_a C_a + C_0^2 \){\one}
  +\(C_a C_b \frac 12 (i f_{abc} + d_{abc})+ (C_0 C_c + C_c C_0)\)\tau_c.
\eeq
If we set $C_0=\frac{-N+3}6$, we have
\beq
C^2 -\frac{(N+1)^2}4 =  \frac 23\(C_a C_a - N(\frac N3+1)\){\bf 1}
      +\frac 12 (F_c + D_c)\; \tau_c
\eeq
for $F_a, D_a$ as defined before, hence
$C_{vac}^2 = \frac{(N+1)^2}4 $ for 
\beq
C_{vac} = \frac{-N+3}6\; +  \xi_a \tau_a.
\eeq
Now consider the action \cite{matrixsphere}
\beqa
S &=&  \frac 1{g D_N} Tr V(C) 
  = \frac 1{g D_N} 
   Tr \((C^2 -\frac{(N+1)^2}4)^2\)  \nn\\
  &=& \frac 1{g} \int \frac 43\; (C_a C^a - (\frac{N^2}3 +N))^2 
    + \frac 12 (F_c + D_c)(F_c + D_c).
\eeqa
As opposed to the 2-dimensional case, 
this does not seem to describe an
interesting gauge theory on fuzzy $\C P^2$, because the auxiliary
fields ``eat up'' the gauge fields: Since for a $C$ with finite action
the eigenvalues and multiplicities of $C$ must be
approximately the same as those of $C_{vac}$,
there exists a basis where $C_a = \xi_a + A_a$ for ``small'' $A_a$.
We can then split $A_a$ into tangential and transversal components as
before, and at the north pole this action is 
\beq
S = \frac 1g \int \frac{16}{9}\; N^2 A_8 A_8  
    + \frac 12 (\sum_{r=1,2,3} (-2N A_r + F_r)^2 +
    (\frac{2N}3 A_8+F_8)^2 + \sum_{t=4}^7 (D_t + F_t)^2)
\eeq
(at the north pole).
Since the $A_a$ are small we can assume that $F_a$ is finite, so that
$A_\mr = o(\frac 1N)$ and $F_\mt = o(\frac 1N)$. 
Hence $A$ is indeed
tangential, and after integrating out the $A_\mr$ one is left with
\beq
S =  \frac 1g \int  \frac 12 \sum_{t=4}^7 (D_t + F_t)^2.
\eeq
The meaning of this is obscure.

\paragraph{Conjugate fields}

The situation becomes somewhat more interesting if one adds a second
``conjugated'' field 
\beq
\tilde C =  \frac{N+6}6\; \one +  C_a \tilde\tau_a
\eeq
where $\tilde{}$ denotes the outer (diagram) automorphism of $su(3)$, 
i.e. essentially $\tilde\tau_a  =\la_a$ are the ordinary Gell-Mann
matrices. Alternatively, the complex conjugate 
$\tau_a^*$ are also equivalent to $\tilde\tau_a$.
Then
\beq
\tilde\tau_a \tilde\tau_b = \frac 23 \d_{ab} 
   + \frac 12(i f_{abc} - d_{abc}) \tilde\tau_c,
\eeq
the only difference to \eq{tau-algebra} being the sign in front of $d_{abc}$.
Then
\beq
\tilde C^2 -\frac{(N+2)^2}4 =  \frac 23\(C_a C_a - N(\frac N3+1)\){\bf 1}
      +\frac 12 (F_c - D_c)\; \tau_c
\eeq
for $F_a, D_a$ as defined before. Therefore 
$\tilde C^2 = \frac{(N+2)^2}4$ for $C_a = \xi_a$, and
we can consider the action
\beqa
\tilde S &=& \frac 1{g D_N} Tr \tilde V(\tilde C) 
  := \frac 1{g D_N} Tr \((\tilde C^2 -\frac{(N+2)^2}4)^2\)  \nn\\
  &=& \frac 1g \int \frac 43\; (C_a C_a - (\frac{N^2}3 +N))^2 
    + \frac 12 (F_c - D_c)(F_c - D_c).
\eeqa
Therefore 
\beq
S + \tilde S =  \frac 1{g D_N} Tr (V(C) + \tilde V(\tilde C))
 = \frac 1g \int \frac 83\; (C_a C_a - (\frac{N^2}3 +N))^2 
    +  F_a F_a + D_a D_a.
\eeq
Now the transversal
$A_\mr$ (at the north pole) decouple due to $D_a D_a$, and one is left with 
essentially $Tr (F_\mr F_\mr + D_\mt D_\mt)$. However, this is still not
what we want because $F_a F_a$ is not covariant in the usual
(Euclidean) sense, and furthermore the meaning of the tangential 
$D_\mt D_\mt$ is not clear.
This is related to the question whether one can solve the constraint
$D_a =0$ exactly in the noncommutative case, or only up to $o(1)$ corrections.

In particular, we make the following observation: for the special
choice $\mu =2$ and $\mu_2 = -\frac 23$, the gauge action \eq{S-gauge-final}
can be written as
\beqa
S_{YM} + S_D  &=& \frac 1{g D_N}\; Tr(\frac 43 (C\cdot C) (C\cdot C) 
    + \frac 12(dCC)(dCC) + \frac 12 (ifCC) (ifCC)   \nn\\
 &&  +2if_{abc} C_aC_bC_c 
  - 4(\frac {2N}3 +1) dCCC + (\frac{4N^2}3 +4N +5) C_a C_a)  \nn\\
 &=&  \frac 1{g D_N}\;  Tr(V_1(C) + V_2(\tilde C))
\label{S-VVp}
\eeqa
for 
\beqa
V_1(C) &=& \frac 12 (C+\frac{N-3}6)^2(C^2-\frac 13(7N+9) C
  +\frac 1{12}(N+3)(11N+25)),  \nn\\
V_2(\tilde C) &=& \frac 12 (\tilde C -\frac{N+6}6)^3 (\tilde C +\frac 52(N+2))
\eeqa
While the meaning of this model is unclear for the same reasons as
above and furthermore  the mass
term for $A_8$ is undetermined (see Section \ref{sec:decoupling}), 
the above form might be accessible to analytical (or numerical) studies. This
certainly motivates further investigations.

\subsection{Quantization}

The quantization of these models is straightforward in principle, by 
a ``path integral'' over the hermitian matrices
\beq
Z[J] = \int d  C_a e^{-(S_{YM} + S_D + Tr C_a J_a)}
\label{ZJ}
\eeq
Note that there is no need to fix the gauge unless one wants to do
perturbation theory, since the gauge orbit is compact. The 
gauge-fixing terms required 
for perturbation theory will not be discussed here, 
cp. \cite{ydri-perturb}. 

We claim that the above path integral is well-defined and finite for 
any fixed $N$ provided $\mu_1 >0$ and $\mu_2 \geq 0$.
To see this, note e.g. by rescaling $C_a \to \a C_a$ that 
it is enough to show that 
$\int d  C_a e^{-Tr ((i[C,C])^2 + \mu_1 (dCC)^2)}$
is convergent, in obvious notation. Using 
$Tr (i[C,C])^2 \geq Tr \frac 1{\sqrt{12}} (ifCC)^2$ and \eq{YM-rewritten},
it is enough to show that 
$\int d  C_a e^{-Tr (C \cdot C)^2}$ is convergent, which is true.

For the approach with constraints in Section \ref{sec:constraints},
the path integral is over  the constrained configuration space
$C^2 = \frac {(N+1)^2}2$, which is an integral over the compact space 
$U(3D_N)/(U(n_+) \times U(n_-))$. In either case, this finiteness
property allows to study quantum field theory on $\C P^2_N$ in a 
very clean way. In both approaches, 
the nontrivial question of course remains whether
the model is renormalizable, i.e. whether 
there exists a suitable scaling of the coefficients in the action
such that the limit $N \to \infty$ of the quantized model is well-defined.
These issues could be studied using the renormalization group 
methods developed in \cite{wulki}.

\section{Topologically nontrivial solutions}
\label{sec:top-nontriv}

In this section, we will construct some explicit solutions of the 
equations of motion \eq{eom-1}, \eq{eom-additional} with finite action,
which in the classical (large $N$) limit 
become topologically nontrivial solutions such as 
monopoles and instantons.

The idea is to identify the solutions of the  gauge 
theory with certain irreps of the symmetry group $SU(3)$, 
which replaces the Poincare group. Recalling that the ``vacuum'' solution 
$C_a = \xi_a =\pi_{(0,N)}(T_a)$ of our action $S_{YM}  + S_D$
is obtained as irrep $V_{(0,N)}$, it is natural to consider other
representations, such as $C_a = \pi_{\L}(T_a)$ for other irreps
labeled by their highest weight $\L$. If $\L$ is close to $(0,N)$,
it seems plausible that they give rise to nontrivial 
saddle-points of the action.
This idea essentially works, with some modifications which are
necessary in the nonabelian case.

\subsection{Monopoles, or $U(1)$ instantons}

In the classical case, $\C P^2$ admits monopole configurations 
with any integer first Chern number or charge, due to the presence of a
nontrivial 2-sphere which generates $H_2(\C P^2)$.
They have been constructed on fuzzy 
$\C P^2$ as projective modules in \cite{wataCPN}. 
Here we will recover them as solutions 
of the equation of motion.

Consider the Ansatz
\beq
C_a = \a \xi_a^{(M)} = \a \pi_{(0,M)}(T_a)
\label{monopole-ansatz}
\eeq
where $\xi_a^{(M)}$ is the generator of $su(3)$ in the \rep $V_{M\L_2}$,
with
\beq
M = N-m.
\eeq
The gauge field $A_a(x)$ is then
determined by considering $C_a$ as a fluctuation
of the vacuum $\xi_a = \xi_a^{(N)}$, i.e.
\beq
C_a = \xi_a +  \; A_a.
\eeq
This means that one should imagine $C_a$ to be a $D_M \times D_M$
block-matrix embedded in the configuration space of $D_N \times D_N$ 
matrices as in \cite{matrixsphere}:
\beq
C_a = \left(\begin{array}{cc} 
\a \xi_a^{(M)} & 0
\\ 0 & 0\end{array}\right) = \xi_a  + A_i \sigma^i
\label{monopole-matrix}
\eeq
(note that $C_a=0$ is also a solution of the equation of motion with
action 0, provided $\mu_1 =0$).
This is clear as long as $M<N$. However we also want to admit $M>N$,
which is achieved by relaxing the condition that $C_a$ be $D_N \times D_N$
matrices. For example, one could fix $C_a$ to have size $D_N + N$,
which for large $N$ admits all relevant solutions of an abelian gauge theory.
Notice also that the particular embedding above determines the
location of a ``Dirac string'' (singularity of the gauge field). This
embedding is of course arbitrary, and the Dirac string can be moved
using a gauge transformation.

The precise normalization $\a$ is 
determined by the equation of motion and depends on $\mu$, but essentially it 
is determined by the constraint $D_a =o(1)$. 
Using
\beq
[\xi_a^{(M)}, \xi^{(M)}_b] = \frac i2 f^{ab}_c \xi^{(M)}_c, \qquad
%g^{ab} \xi_a^{(M)} t_b    &=& \frac 13 M^2 +M, \label{deft2} 
d^{ab}_c \xi_a^{(M)} \xi^{(M)}_b = (\frac{2M}3 +1) \; \xi^{(M)}_c  
\label{t-relations}
\eeq
one finds that $D_a=o(1)$ implies 
\beq
\a = 1 + \frac mN + o(1/N^2)
\label{alpha-mono}
\eeq
where the $o(1/N^2)$ term depends on $\mu$. This also
implies $C_a C_a = \frac 13 N^2+N+ o(1)$, hence
\eq{monopole-ansatz} describes 
indeed a tangential gauge field on fuzzy $\C P^2$.
The gauge field is then guaranteed to be tangential \eq{F-tang}, 
and is given by
\beq
F_{ab} =  \frac 12\; f_{ab}^c (-\a^2 + \a) \xi^{(M)}_c
       \approx - \frac m2\; f_{ab}^c  \frac{\xi_c^{(M)}}N.
\eeq
$A_a$ turns out to be finite except at the ``south sphere'', and will be 
calculated explicitly below.
This implies that 
\beq
\frac{\xi_a^{(M)}}N = \frac{\xi_a}N +o(1/N) \approx \frac{x_a}{ \sqrt{3}},
\eeq
hence 
\beq
F_{ab} = -\frac m{2\sqrt{3} }\; f_{ab}^c x_c, 
\label{F-mono-comp}
\eeq
or
\beq
 = -2i\pi m\; \omega
\eeq
is a multiple of the symplectic form $\eta = 3 \pi \omega$ \eq{omega}. 
In particular, $F$ is selfdual.
Using the formulas in Section \ref{subsec:chern}, the first 
Chern number in the large $N$ limit is
\beq
c_1 =  m.
\eeq
This shows that $A_a$ is indeed a connection on a bundle, and
we have recovered the result of \cite{wataCPN} in a somewhat different
formulation. 
Such monopoles do not exist on $S^4$.
The value of the action is 
\beq
S_{YM} = \frac 1g \int F_{ab} F_{ ab} 
 = \frac 1g \int 12 m^2  
\eeq
plus corrections of order $o(1/N)$, and $S_D$ does not contribute
for large $N$.

\subsubsection{Explicit form of the gauge field}

We want to calculate the gauge field 
\beq
 \; A^{mono}_a(x) = \a \xi_a^{(M)} - \xi_a 
\eeq
explicitly.
We fix the gauge (using a $U(D_M)$ gauge transformation) 
by diagonalizing both $\xi^{(M)}_3$ and $\xi^{(M)}_8$, and 
matching the eigenvalues with those of $\xi_{3,8}$ except for 
the lowest eigenvalues of $\xi_8$. This amounts 
to a embedding of the corresponding matrices, and putting the 
singularity at the south sphere. 
Geometrically, it amounts to match the ``upper parts'' 
of the weight triangles of the \reps 
$V_{M \L_2}$ and $V_{N \L_2}$. 
Since all multiplicities are one, one could simply work with the
$su(2)$ subalgebras. However as a warm-up 
for the instanton calculation, we shall use the
Gelfand-Tsetlin basis, where the operators 
can be calculated explicitly. 
One finds that (see \eq{monopole-gens} in Appendix F)
the operators $\xi_a^{(M)}$ can be written in terms of the 
$\xi$ as 
\beqa
\xi^{(M)}_{1,2,3} &=& \xi_{1,2,3}, \qquad 
      \xi^{(M)}_{8} = \xi_{8} - \frac m{\sqrt{3}}, \nn\\
\xi^{(M)}_4 \pm i \xi^{(M)}_5 &=& 
     (\xi_4 \pm i \xi_5)
   -m\frac{\sqrt{3}}{2}  \frac{x_4 \pm i x_5}{2 x_8+ 1},\nn\\
\xi^{(M)}_6 \pm i \xi^{(M)}_7 &=& 
   (\xi_6 \pm i \xi_7)
   - m\frac{\sqrt{3}}{2} \frac{x_6 \pm i x_7}{2 x_8+1}
\label{monopole-generators-explicit}
\eeqa
using  $x_a = \frac{\sqrt{3}}N \xi_a$.
There are further correction terms of order $\frac{1}{N(2x_8+1)^2} x$,
which vanish for large $N$ as long as $2 x_8+1 >0$, i.e. away
from the south sphere. 
We can now find $\a_m$ by requiring that $A_8=0$ at the north
pole, which using $\xi_8|_{NP} = \frac N{\sqrt{3}}$ gives
\beq
\a = \frac N M = 1+\frac mN + o(1/N^2)
\eeq
in agreement with \eq{alpha-mono}. Then the gauge field for large $N$ is
\beqa
A_{1,2,3}^{mono} &=& \frac m{\sqrt{3}}\; x_{1,2,3}, \qquad  
    A_8^{mono}  = \frac m{\sqrt{3}}\;(x_8-1), \nn\\
A_4^{mono} \pm i A_5 ^{mono} &=& 
    \frac m{\sqrt{3}}\;(1-\frac 3{2} \frac 1{2 x_8+1})
                        (x_4 \pm i x_5),\nn\\
A_6^{mono} \pm i A_7^{mono}  &=& 
      \frac m{\sqrt{3}}\;(1-\frac 3{2} \frac 1{2 x_8+1})
                        (x_6 \pm i x_7).\nn
\eeqa
In the classical limit,
these formulas are valid everywhere except  
at the lowest eigenvalue of $2x_8 = - 1$, i.e. the south
sphere.  One can check easily that $x_a A_a = 0$.

\subsection{Nonabelian case: $U(2)$ instantons}
\label{sec:instantons}

We will exhibit here nontrivial solutions of the equation of motion 
\eq{eom-1}
for the nonabelian $U(2)$ gauge theory defined by the action 
\eq{S-gauge-final},
for matrices $C_a$ of size $ \approx 2 D_N$. In 
the classical limit, they
describe $U(2)$ gauge fields with nontrivial first and second Chern
number and finite action.

We want to generalize the above construction 
for other representations such as $V_{(1,N)}$. Here 
some modification is required,
which can be understood as follows: according to \cite{qfuzzybranes},
$C_a' = \pi_{(1,N)}(T_a)$ can be
considered as quantization of a 6-dimensional adjoint orbit
infinitesimally close to $\C P^2_N$, which can (heuristically) be viewed as 
``bundle'' over $\C P^2_N$ whose fiber is a non-commutative 
2-point space.
This $C_a'$ does not satisfy the constraint $D_a =0$ of $\C P^2$,
and we have to modify $C_a'$; hence the prime. 
This can be done using the map $\cD^{nl}$ defined in 
Section \ref{sec:nc-constraints}, which leads to
an instanton-like solution. There is a small caveat: our
construction will only give us the uniform, non-localized
instanton with instanton number $1$, 
and not the full moduli space of instantons \cite{buchdahl}. 
One may hope that a modification of the
construction presented here will also give localized instantons.
Furthermore, our solutions are not (anti)selfdual since they also 
contain certain $U(1)$ monopoles.

Even though we do not attempt it here, it seems plausible that
the generalization of this construction to $V_{(k,N)}$ leads to 
solutions with instanton number $k$.

\subsubsection{Group-theoretical origin}

Before exhibiting an exact solution  of our model, 
we give an approximate derivation which is valid for large $N$.
From a group-theoretical point of view,
there is a natural candidate for a nontrivial saddle point
 generalizing \eq{monopole-ansatz} given by the Ansatz 
\beq
C_a'' =  \zeta_a^{(m)} =  \pi_{M\L_2+\L_1}(T_a)
\label{instanton-ansatz}
\eeq
where $\zeta_a^{(m)}$ 
is the \rep with highest weight $M\L_2 + \L_1$, 
and 
$$M = N-m. 
$$
Since 
$\dim(V_{N\L_2+\L_1}) \approx 2\; \dim(V_{N\L_2})$, 
we can hope to embed $V_{M\L_2+\L_1}$ in $V_{N \L_2} \tens \C^2$, defining a
gauge field by
\beq
A_a'' = \zeta_a^{(m)} - \xi_a. 
\eeq
In fact, we will see below that $A_a''$ is finite
and well-defined in the classical limit (except at the south sphere).
Unfortunately, the above Ansatz is not admissible
because $A_\mr''$  is not tangential.
However, this can be fixed in a $su(3)$-covariant 
and gauge invariant way, 
by subtracting the transversal components
using the tensor maps $\cD^{lin}$ or $\cD^{nl}$. 
Recall that the linear map $\cD^{lin}$
separates the tangential from the transversal components, and splits
the latter into two subspaces: one component parallel to
$\xi_a$ (i.e. $A_8''$ at the north pole) 
with approximate eigenvalue $+ \frac 13$, and the complement 
(i.e. $A_{1,2,3}''$ at the north pole) with approximate eigenvalue 
$-1$. The first can be absorbed simply by redefining 
$$
C_a' = \a' \zeta_a^{(m)} = \xi_a + A_a'
$$  
such that $C'\cdot C' = \xi\cdot \xi$ and
hence $x \cdot A' =0$. This leads to
\beq
\a' = 1+\frac{2m-1}{2N} +o(1/N^2).
\eeq
Now the remaining non-tangential components have eigenvalue
$-1$ of $\cD^{lin}$, and hence can be subtracted by adding
$\cD^{lin}$. Since $\cD^{lin}$ acting on $C$ coincides with
$\cD^{nl}$, this
leads to the modified Ansatz
\beqa
C_a &=& C_a' + \frac{1}{2N}(d_{abc} C_b' C_c' -(\frac{2N}3+1) C_a') \nn\\
  &=& \a \zeta_a^{(m)} + \b(d_{abc} \zeta_b^{(m)} \zeta_c^{(m)} -(\frac{2M+7}3)
  \zeta_a^{(m)})) +  o(1/N^2)\;\zeta_a^{(m)}
\label{instanton-proj}
\eeqa
where the last equality holds for
\beq
\a = 1+\frac m N, \qquad
\b = \frac{\a^2}{2N} = \frac 1{2N} + \frac{2m-1}{2N^2} + o(1/N^3).
\label{a-b-ansatz}
\eeq
Therefore the gauge fields $A$ defined by
\beq
C_a = \xi_a + A_a
\eeq
are finite, and satisfy $D^{lin}(A)= o(1/N)$. 
Hence the non-tangential components of 
$A$ are of order $1/N$, or equivalently
$dCC -(\frac {2N}3 +1) C = o(1)$ and
$C \cdot C - \frac{N^2}3 +N =o(1)$. 

To summarize, we define the (approximately tangential) gauge fields 
by\footnote{the particular form of $\tilde D^{(m)}$ will  be 
understood later}
\beq
C_a = \a \zeta_a^{(m)} + \b \tilde D^{(m)}_a = \xi_a +  A_a 
\label{inst-ansatz-2}
\eeq
where
\beq
\tilde D^{(m)}_a = d_a^{bc} \zeta^{(m)}_b \zeta^{(m)}_c 
    - \frac 13(2M +7) \zeta^{(m)}_a.
\eeq
In fact $\zeta_a^{(m)}$ and $\tilde D^{(m)}_a$ are the only two 
tensor operators on $V_{N\L_2+\L_1}$ with an index in the adjoint.
This strongly suggests 
that there should be an exact solution  for the above 
Ansatz \eq{inst-ansatz-2}. This is indeed the case, 
and the result derived in Section \ref{sec:solution-exact} 
is in agreement with this approximate derivation.
But first, we compute the corresponding field strength for large $N$.

\subsubsection{Field strength, action and Chern class}

Before showing that 
\eq{inst-ansatz-2} indeed contains an exact solution of our model, 
we can get some insight by evaluating the field
strength approximately at the north pole. 

Since the gauge fields and field strength are operators on $V_{N\L_2+\L_1}$, 
there is a natural action of $su(3)$ on gauge
fields, generated by the vector field resp. Lie derivative
\beq
\cL_a = [\zeta_a, .] 
\eeq
It corresponds to a $SU(3)$ Killing vector 
field\footnote{On scalar fields (i.e. $\phi \in Mat(V_{N\L_2+\L_1})$), it 
coincides with the usual Killing vector field
$\cL'_a \phi = [\xi_a, \phi]$ in the large $N$ limit,
because $[A_a, f(x)]$ tends to zero
as long as we stay away from the ``south sphere''.}
in the classical limit.
We extend its action to tensor fields with indices in $su(3)$ in the 
natural way (cp. \cite{paolo}), in particular 
\beq
\cL_a C_b = [\zeta_a, C_b]  - \frac i2 f_{abc} C_c
\eeq
and similarly for higher-rank tensor fields. It satisfies the $su(3)$
algebra
\beq
[\cL_a,\cL_b] = \frac i2 f_{abc} \cL_c.
\eeq
In particular, the field strength transforms as
\beq
\cL_a F_{bc} = [\zeta_a, F_{bc}] - \frac i2(f_{abd} F_{dc}  + f_{acd} F_{bd}).
\eeq
This can be used to rotate any ``point'' to the north pole, 
and reduce the computation of $F_{ab}(x)$
to a calculation it at the north pole. 
Now if $C_a$ is the above instanton 
\eq{inst-ansatz-2}, covariance \eq{D-covariance} implies that both
the instanton field $C_a$ as well as the corresponding field strength 
are {\em invariant} under this action of $SU(3)$.
Therefore $F_{ab}$ is constant apart from a rotation of the indices.

Let us therefore evaluate $F_{ab}$ at the north pole.
In view of the first line of \eq{instanton-proj}, 
we have $C_a = \a' \zeta_a^{(m)} + \cD^{lin}(A')_a + o(1/N) = \xi_a + A_a$. 
Therefore
\beq
C_\mt \approx \a' \zeta^{(m)}_\mt, \quad C_\mr \approx \xi_\mr
\eeq
($\approx$ stands for equal up to $o(1/N)$)
at the north pole, because the correction term $\cD^{lin}(A')$ 
is transversal. This is also sufficient to calculate commutators
at the north pole, as long as the fields are smooth for large $N$.
Then using $\a' \zeta^{(m)}_\mr = \xi_\mr +  A_\mr'$ we get
\beqa
F_{\mt \mt} &\approx& \a'^2 i[\zeta^{(m)}_\mt,\zeta^{(m)}_\mt] 
        + \frac 12 f_{\mt\mt\mr} C_\mr 
      \approx \frac 12 f_{\mt\mt\mr} (-\a'^2 \zeta^{(m)}_\mr + \xi_\mr) 
  = \frac 12 f_{\mt\mt\mr} (\xi_\mr(1-\a') - \a'  A_\mr') \nn\\
   &\approx& \frac 12 f_{\mt\mt\mr}(\frac{\frac 12-m}{\sqrt{3}} x_\mr - \a' A_\mr') \label{F-inst-tt}\\
  %  \approx \frac 12 f_{\mt\mt\mr} ((-\a'^2+\a') t_\mr) \nn\\    
F_{\mt\mr} &\approx& \a' i [\zeta^{(m)}_\mt,C_\mr] 
             - \frac 12 f_{\mt\mr\mt} C_\mt   
       = \frac 12 f_{\mt\mr\mt} (\a' C_\mt - C_\mt) 
   =\frac 12  f_{\mt\mr\mt} (\a'-1) (\xi_\mt + A_\mt) 
    \approx \frac 12 f_{\mt\mr\mt} x_\mt  \approx 0 \nn\\
F_{\mr\mr} &\approx& i[\xi_\mr,\xi_\mr] - \frac 12 f_{\mr\mr\mt}\xi_\mt = 0 
\nn
\eeqa
This shows explicitly that $F$ is purely tangential, 
since $F_{tr} \approx x_t=0$ at the north pole.
This was shown in general in Section \ref{sec:constfield}.
Using the explicit results \eq{instantonA-1} (which does not require
the full calculation in Appendix E), 
the tangential components are
\beqa
F_{\mt \mt}  &\approx&
 f_{\mt\mt 8}\frac{1}{2\sqrt{3}} \frac{1-2m}2 \one
  - \frac 1{4} (f_{\mt\mt 3} \s_3 + f_{\mt\mt 1} \s_1 + f_{\mt\mt 2} \s_2)\nn\\
  &=& F_{\mt \mt}^{U(1)} \one + F_{\mt \mt}^{SU(2)}
\label{F-inst-NP}
\eeqa
at the north pole, which extends to the entire $\C P^2$ since
the field strength is invariant under $\cL_a$. 
Comparing with the results of Section \ref{subsec:selfdual-class}, 
we see that $F_{tt}$ can be split into a selfdual part
proportional to the symplectic form $\eta$
(this is the monopole part), and an anti-selfdual part; the latter is
identified as instanton. 
Note that there is no $m$ where the field
strength is purely selfdual or purely antiselfdual. Therefore this 
solution is somewhat different from the usual instantons. 
It is plausible that the $U(1)$ monopole part will play an important
role in the coupling to fermions, due to the spin${}^c$ structure of
$\C P^2$ \cite{hawking}.

Hence we have a gauge field which 
is finite and well-defined in the commutative limit 
(apart from a null-set on the south sphere, see below),
and its field strength is globally finite. To see that this really 
describes a connection on some bundle on $\C P^2$, 
we need to check that the first and second  Chern numbers are integer. The
corresponding  $U(2)$ bundle structure will then be determined.
Using
\beq 
tr_{2\times 2} F_{ab} = (2m-1) F_{ab}^{mono}
\eeq
and denoting the basic 2-cocycle with $\omega = \frac{\eta}{3\pi}$,
we can read off the first Chern class resp. number as
\beq
c_1 = (2m-1) \omega \;\;\cong \; 2m-1.
\label{c1-nonabel}
\eeq
As a check, one can also compute $c_1$ directly from \eq{c1-4dim},
without using the invariance under $SU(3)$. Using
\beq
\frac 1N F_a C_a = \frac{-2m+1}2
\label{FC-eq}
\eeq
for large $N$ (see also Section \ref{sec:solution-exact}), 
we find again \eq{c1-nonabel}.
Similarly, the 2nd Chern class is
\beqa
c_2 &=& -\frac 1{8\pi^2} ( tr F \wedge tr F - tr(F\wedge F)) \nn\\
  &=& -\frac 1{8\pi^2} \(  tr F^{U(1)} \wedge tr F^{U(1)}
   - tr((F^{U(1)}+ F^{SU(2)}) \wedge(\star F^{U(1)}- \star F^{SU(2)}))\) \nn\\
  &=& -\frac 1{8\pi^2}   tr((F^{U(1)}+ F^{SU(2)}) \wedge
        \star (F^{U(1)}+ F^{SU(2)})) \nn\\
 &=&  \frac 12 tr( F_{ab} F_{ab}) (\frac 2{9\pi^2} d^4 V) 
     =  \frac 12 tr( F_{ab} F_{ab}) \; \omega^2 
\eeqa
using \eq{2forms-inner}, \eq{cycle-normalization}, \eq{F-form} and 
and the fact that $F^{SU(2)}$ is anti-selfdual while $F^{mono}$
is selfdual. Curiously, we find that $c_2$ coincides with the 
Yang-Mills action, even though $F$ is neither selfdual nor
anti-selfdual.
This may hint at some underlying BPS structure. 

Using \eq{F-inst-NP}, we find
\beq
\frac 12\; tr_{2\times 2} F_{ab} F_{ab} = 1-m+m^2.
\label{YM-direct}
\eeq
This agrees with the direct calculation of the Yang-Mills action, 
using \eq{YM-rewritten1}, as shown
in \eq{YM-value-2}.
Therefore
\beq
c_2 = 1-m+m^2.
\eeq
Note that since $c_1  =2m-1$ is obtained by taking the trace over a
$2\times 2$ matrix, the individual $U(1)$ charges are 
half-integer, $m-\frac 12$. 
This is consistent with the explicit form of the gauge potential
given in \eq{instantonA-1}, which splits into 
$ \frac 12 (-2m+1) \one_{2\times 2}$ plus a $SU(2)$ part.
A similar observation is exploited in 
\cite{dolan1,dolan2} to construct the charges of the standard model
from such nontrivial bundles. 
This is quite remarkable in view
of the common belief that (non-expanded) noncommutative gauge theory 
 allows only the charges $\pm 1, 0$: if we succeed to couple
the gauge field to a fermion, this may
show a way to circumvent this restriction without expanding in a
deformation parameter and using the Seiberg-Witten map \cite{wess}.

The explicit form of the instanton gauge field can now be worked out 
using the Gelfand-Tsetlin basis. Because it is somewhat lengthy, we 
give the result in Appendix E.
These field configurations 
correspond to (generalized) instantons which are ``spread out'' over
the entire $\C P^2$, since $F_{ab} F_{ab} =const$. It would be very 
interesting to find localized versions of these instantons. 
In particular,
one would like to have ``many-instanton'' solutions, and a description
of the corresponding moduli spaces.
Furthermore, it would also be desirable to understand better the 
connection with the usual self-dual instantons, and / or
to construct fuzzy analogs of those for finite $N$.

\subsubsection{The classical bundle structure}
\label{sec:class-bundle}

Let us compute the total Chern character
for the above field configuration:
\beq
ch = 2+c_1+\frac 12(c_1\wedge c_1-2c_2) 
 = 2+  (2m-1) \omega + (m^2-m-\frac 12) \omega^2.
\eeq
As shown in \cite{dolan1,dolan2}, 
this is precisely the Chern character
of the bundle 
\beq
L^{m} \tens F
\eeq
where $L$ is the generating line bundle (i.e. the above monopole bundle with
charge $m=-1$) 
and $F$ is a nontrivial rank 2 bundle over $\C P^2$ with structure
group $U(2)$ defined by $F \oplus L  \cong I^3$, where $I^3$ is the
trivial rank 3 bundle. Note that $c_1(F) = -c_1(L)$ and
$c_2(F) = c_1(L)^2 \cong 1$.

Therefore our ``instanton'' gauge field $A_a$ is a connection on the
nontrivial $U(2)$ bundle $L^m \tens F$. 
It is a solution of $U(2)$ Yang-Mills with
finite action, which is neither selfdual nor
anti-selfdual. These bundles are
associated to the principal $U(2)$ bundle 
\beq
\matrix{U(2)&\longrightarrow & SU(3) \cr && \downarrow \cr &&\  \C P^2.\ \cr}
\eeq
As explained in \cite{stroh,dolan1,dolan2}, a Dirac operator
can be defined on this (classical) bundle taking into account the 
spin${}^c$ structure,
for spinors transforming under $SU(2) \times U(1)$.
All this strongly
suggests that it should be possible to define such a Dirac operator also 
in the fuzzy case in our matrix 
approach\footnote{see \cite{bala,stroh,coset} for possible
definitions of a Dirac operator on fuzzy $\C P^2$}, and that
physically interesting models 
including matter could be constructed as matrix models on
fuzzy $\C P^2_N$.

We can also see directly how our construction of fuzzy 
instantons leads in the classical limit to the bundle $L^m \tens F$. 
According to the discussion around \eq{phi-action}, 
the (fuzzy version of)
sections of the associated $U(2)$ vector bundle on which 
$C$ acts from the left are given by rectangular matrices
$Hom(V_{(0,N)}, V_{(1,M)})$. The harmonic analysis of this space of ``fuzzy'' 
sections, i.e. their decomposition as $SU(3)$ module is therefore
\beqa
V_{(0,N)}^* \tens V_{(1,M)} &=& \bigoplus_{k=0}^{M} 
  (V_{(k,k+m+1)} \oplus V_{(k,k+m-1)}) ,\qquad m \geq 1   \nn\\
V_{(0,N)}^* \tens V_{(1,M)} &=& \bigoplus_{k=0}^{N} 
  (V_{(k-m,k+1)} \oplus V_{(k-m,k)}) ,\qquad m < 1  
\eeqa
This coincides precisely with the more formal constructions in
\cite{stroh}, where it was also shown that
it agrees with the space of sections in the bundle\footnote{$F$
is denoted with $H^{-1}$ in  \cite{stroh}} $L^m \tens F$ as $N \to\infty$.

\subsection{Solution of the equation of motion}
\label{sec:solution-exact}

In order to show that \eq{inst-ansatz-2} contains exact solutions of
the  equation
of motion, we must 
learn to work with the $\zeta_a = \zeta_a^{(m)}$ operators, which 
act on $V_\L$ with $\L = M \L_2 + \L_1$. 
The following combination 
turns out to be useful:
\beq
\tilde D:= \tilde D^{(m)} = (\frac{-M+1}6 + X)^2 - (\frac{M+1}2)^2 
= \frac 13\(M+ 2\) +\frac{1}2 \tilde D_c\; \tau^c.
\label{D1-def}
\eeq
Here
\beq
\tilde D_c = {d^{ab}}_c \zeta_a \zeta_b - \frac 13(2M +7) \zeta_c,
\label{D1m-def}
\eeq
$X = \zeta_a \tau^a$, and
$f \zeta \zeta + 3\zeta =0$ as well as \eq{instanton-casimir} have been used.
This is the combination which occurs in \eq{instanton-proj}.
Using \eq{char-X}, it follows that $\tilde D$ is a projector on
$V_{\L-\L_1}$, and it vanishes on $V_{\L+\L_2}$ and
$V_{\L+\L_2-\a_2}$; this is the origin of the above definition.
The $\tilde D_a$ are clearly covariant, i.e. 
they transform in the 8 of $su(3)$: 
\beq
[\zeta_a,\tilde D_b] = i/2 f_{ab}^c \tilde D_c
\label{D-covariance}
\eeq
hence they are ``tensor operators''.
Using these properties, 
One can derive (see Appendix D) the following
identities which hold on $V_\L$:
\beqa
\zeta_a \zeta_a &=& \frac 13 (M+2)^2   \label{instanton-casimir}  \\
\tilde D_a \zeta_a &=& -\frac{(M+2)(M+8)}{3}  \nn\\
\tilde D_a \tilde D_a &=& \frac 13 (M+2)(10M+32)  \nn\\
d^{ab}_c \zeta_a \zeta_b &=&  \tilde D_c +\frac 13 (2M+7) \zeta_c  \nn\\
i f^{ab}_c \tilde D_a \zeta_b &=& -3 D_c   
   = i f^{ab}_c  \zeta_a \tilde D_b \nn\\
i f^{ab}_c \tilde D_a \tilde D_b &=&  4(M+2) \zeta_c + 2(2M+7) \tilde D_c \nn\\
d^{ab}_c \tilde D_a \zeta_b &=& -\frac 43(2 + M) \zeta_c 
  -\frac 13(2M+7) \tilde D_c = d^{ab}_c \zeta_a \tilde D_b   \nn\\
d^{ab}_c \tilde D_a \tilde D_b &=& -4(M+2) \zeta_c  
+ \frac 23(2M+7) \tilde D_c .
\eeqa
It is then easy to see that the Ansatz 
\beq
C_a = \a \zeta_a + \b \tilde D_a
\label{C-inst-ansatz}
\eeq
contains an exact solution of the equation of motion \eq{eom-1} for finite
$N$ with $\a,\b$ as specified in \eq{a-b-ansatz}. This is so because
using the above identities,
the equation of motion \eq{eom-1} takes the form 
\beq
r(\a,\b) t_a + s(\a,\b) \tilde D_a =0
\eeq
for certain  functions $r,s$
which depend on the coefficients $\mu_{1,2}$ in
the action. To find the solutions to $r=s=0$ we just have to minimize
the positive definite action for $C$ of the form \eq{C-inst-ansatz}, 
which is possible.
Indeed one can easily see that for
\beq
\a = 1+\frac m N +  o(1/N^2), \qquad
\b = \frac 1{2N} +  o(1/N^2);
\label{ab-cond}
\eeq
the constraint 
$D_a = o(\frac 1N) \zeta_a + o(\frac 1N) \tilde D_a =o(1)$ 
and also $C_a C_a - (\frac{N^2}3 +N) = o(1)$ as $N \to \infty$. 
This is sufficient to guarantee that the gauge fields are tangential
according to \eq{F-tang}, and that the constraint term vanishes $S_D \to 0$ 
for $\mu_{1,2} = o(1/N)$.
Plugging this into \eq{YM-rewritten1} gives the Yang-Mills action,
\beq
S_{YM}  = \frac 1g \int tr(1-m+m^2) \quad + o(1/N)
\label{YM-value-2}
\eeq
in complete agreement with the direct computation \eq{YM-direct}.
Here $tr$ is the trace over the nonabelian matrix content, i.e.
$tr_{2\times 2}$ for the $U(2)$ case. Similarly one verifies
\eq{FC-eq}.

One can furthermore show that
the constraint $D_a = 0$
does have exact solutions for the above Ansatz \eq{C-inst-ansatz}
consistent with \eq{ab-cond}, but there exists no solution for
$fCC \propto C$ and $dCC \propto C$ simultaneously. 
This leads to the problem that while the Ansatz is compatible with
the constraint $C^2 = (\frac{N+1}2)^2$
discussed in Section \ref{sec:constraints}, it is no longer a solution
of the corresponding equations of motion. This is
the reason why we concentrate on the formulation with auxiliary
variables.

\section{Discussion and outlook}

We give in this paper a matrix-model formulation of gauge theory
on fuzzy $\C P^2$. Our action differs from related matrix models 
in the context of string theory \cite{ikkt,ARS} by adding
Casimir-type constraint terms following \cite{matrixsphere}, 
which are designed so as to stabilize 
the fuzzy $\C P^2$. From a field theoretic point of view, they 
give the non-tangential degrees of freedom a 
large mass. This ensures that the usual Yang-Mills action is
reproduced in the commutative (large $N$) limit.
We then proceed to find nontrivial solutions of the equation of motion,
which turn out to be $U(1)$ monopoles and certain $U(2)$  
instanton-like solutions. 
 
The main merits of these models are that the quantization is
well-defined and finite, and that topologically nontrivial
configurations arise simply as solutions of the matrix equations of motion. 
In particular, we do not have to sum over 
disconnected topological sectors; they are included in the 
``path'' integral over all matrices. Unlike in 2 dimensions 
\cite{matrixsphere}, it would be too much to expect that
the model can be solved analytically. However, one may hope that
the formulation as matrix model will give a
new handle on 4-dimensional gauge theory. We find one 
interesting simplification at a particular (if unphysical) point in 
parameter space \eq{S-VVp}, which might be interesting to pursue.

There are many open issued which deserve further investigations. 
One is the inclusion of fermions, which is nontrivial due to the 
fact that $\C P^2$ has no spin but spin${}^c$ structure. There
are several papers where this is investigated \cite{bala,stroh,coset},
but the appropriate
coupling to a gauge field in our formulation is not clear. 
Another open problem is to find ``localized'' instantons
and their moduli space. This is complicated by the apparent lack of a
Hodge-star (with correct classical limit) on $\C P^2_N$.
In particular, our
instantons contain a nontrivial $U(1)$ sector, and are neither
selfdual nor anti-selfdual. The $U(1)$ monopole part seems to be related to
the spin${}^c$ structure on $\C P^2$, and may be important for the
coupling to fermions \cite{hawking}.

We also give an alternative formulation by imposing the
constraint \eq{C2-const} rather than using auxiliary fields. The relation of
these different formulations is not clear, in particular we have not found
instanton solutions in the constrained case.
Finally, it would be very desirable to get some insight into the large
$N$ behavior of the quantized model. This could be studied using 
renormalization group techniques developed in \cite{wulki}.

\vskip 10pt

{\bf \large Acknowledgements} 

\vskip 10pt

We would like to thank Paolo Aschieri, Brian Dolan, Branislav Jurco,
John Madore, Helmuth Urbantke, and Ursula and Satoshi Watamura 
for useful discussions on various aspects of this work. In particular,
we want to thank Julius Wess for his support and encouragement. H.S.
is very grateful for invitations to the Erwin-Schr\"odinger
Institute in Vienna where part of this research was carries out,
and H.G.  for invitations to the Ludwig-Maximilians
Universit\"at and the
Max-Planck-Institut f\"ur Physik in Munich.

\section*{Appendix A: Explicit form of the $su(3)$ generators}
\addcontentsline{toc}{subsection}{Appendix A: Explicit form of $su(3)$
  generators}

Denote the simple roots of $su(3)$ with $\a_1, \a_2$, and the
highest root with $\theta = \a_1+\a_2$.
The corresponding Cartan-Weyl generators satisfy
the usual relations 
\beqa
\;[X^+_i,X^-_j ] &=& \d_{i,j} H_{\a_i}, \qquad i=1,2 \nn\\
\;[X_1^+,X_2^+ ] &=& X_\theta^+ \nn
\eeqa
etc. 
Furthermore, 
$[X_\theta^+,X_2^-] = X_1^+, \; [X_2^+,X_\theta^-] = X_1^-$.
It is also useful to define 
\beqa
 H_3 &=& H_{\a_1}, \nn\\
H_\theta &=& H_{\a_1} + H_{\a_2},\nn\\
\sqrt{3}\; H_8 &=& H_{\a_1} + 2 H_{\a_2}.
\eeqa
This provides a normalized basis $T_a, \; a=1,2,...,8$ 
of $su(3)$ defined by
\beqa
 T_4 \pm iT_5 &=& X_\theta^\pm,  \qquad
T_6 \pm i T_7 =  X_2^\pm, \qquad
T_1 \pm i T_2 = X_1^\pm, \nn\\
T_3 &=& \frac 12 H_3, \qquad  T_8 = \frac 12 H_8.
\label{T-generators}
\eeqa
The usual Gell-Mann matrices are then given by 
\beq
\la_a =  2\pi_{\L_1}(T_a)
\label{la-def}
\eeq
where $\pi_{\L_1}$ denotes the defining representation $V_{\L_1}$
with weights
$\nu_i = (\L_1,\L_1 - \a_1,\L_1 - \a_1 - \a_2 = -\L_2)$.
However, for our purpose it is convenient to use the 
``conjugated'' Gell-mann matrices, defined by
\beq
\tau_a =  2\pi_{\L_2}(T_a).
\label{tau-def}
\eeq
Here the weights of the dual  representation $V_{\L_2}$ are
$\nu_i = (\L_2,\L_2 - \a_2, -\L_1)$.
They are explicitly given by 
\beqa
\tau_1 &=&  \(\begin{array}{ccc}0 & 0 & 0\\
                               0 & 0 & 1 \\
                               0 & 1 & 0 \end{array}\),\qquad
\tau_2 =  \(\begin{array}{ccc}  0 & 0 & 0\\
                               0 & 0 & -i \\
                               0 & i & 0 \end{array}\),\qquad
\tau_3 =  \(\begin{array}{ccc} 0 & 0 & 0\\
                              0 & 1 & 0 \\
                              0 & 0 & -1 \end{array}\),\nn\\
\tau_4 &=&  \(\begin{array}{ccc}0 & 0 & -1\\
                               0 & 0 & 0 \\
                               -1 & 0 & 0 \end{array}\),\qquad
\tau_5 =  \(\begin{array}{ccc}0 & 0 & i\\
                               0 & 0 & 0 \\
                               -i & 0 & 0 \end{array}\),\nn\\
\tau_6 &=& \(\begin{array}{ccc} 0 & 1 & 0\\
                                1 & 0 & 0 \\
                                0 & 0 & 0 \end{array}\),\qquad
\tau_7 =  \(\begin{array}{ccc}  0 & -i & 0\\
                               i & 0 & 0 \\
                               0 & 0 & 0 \end{array}\),\nn\\
\tau_8 &=&  \frac 1{\sqrt{3}}\(\begin{array}{ccc} 2 & 0& 0\\
                             0 & -1 & 0 \\
                             0 & 0 & -1 \end{array}\).
\eeqa
One can now compute the explicit form of $f_{abc}$ and $d_{abc}$ 
defined by \eq{tau-algebra}:
\beqa
{f_{12}}^3 &=& 2, \; {f_{14}}^7 = 1,\; {f_{15}}^6 = - 1,\; 
  {f_{24}}^6 = 1, \; {f_{25}}^7 = 1, \nn\\
{f_{34}}^5 &=& 1, \; {f_{36}}^7 = - 1,\; 
{f_{45}}^8 = \sqrt{3}, \; {f_{67}}^8 = \sqrt{3},
\label{f-explicit}
\eeqa
and
\beqa
{d_{11}}^8 &=& -2/\sqrt{3}, \; {d_{14}}^6 = -1,\; {d_{15}}^7 = -1,\; 
{d_{22}}^8= -2/\sqrt{3}, \; {d_{24}}^7 = 1,\; {d_{25}}^6 = -1\nn\\
{d_{33}}^8 &=& -2/\sqrt{3}, \; {d_{34}}^4 = -1,\; \; {d_{35}}^5 = -1,\; 
  {d_{36}}^6 = 1,\; {d_{37}}^7 = 1, {d_{44}}^8 = 1/\sqrt{3}, \nn\\
 {d_{55}}^8 &=& 1/\sqrt{3},\;  
 {d_{66}}^8 = 1/\sqrt{3}, \; {d_{77}}^8 = 1/\sqrt{3}, \; 
{d_{88}}^8 = 2/\sqrt{3}.
\label{d-explicit}
\eeqa
Note also that
\beq
\sum_{a,b} f_{abc} f_{abd} = 12 \d_{cd}.
\label{ff-id}
\eeq

\section*{Appendix B: characteristic equation}
\addcontentsline{toc}{subsection}{Appendix B: characteristic equation}

Consider
\beq
X = \sum_a t_a \tau^a = \frac 12 ((t_a + \tau_a)^2 - t_a t^a - \tau_a \tau^a)
\eeq
and recall the eigenvalues of the quadratic Casimirs 
on the highest weight \rep $V_{N_1 \L_1 + N_2 \L_2}$ are given by
\beq
c_2(\L) = (\L, \L+2\rho) = \frac 23(N_1^2 + N_2^2 + N_1 N_2) +2(N_1+N_2)
\eeq
where $\rho = \L_1 + \L_2$ 
is the Weyl vector of $su(3)$. 
We used here
\beq
(\L_1,\L_1) = \frac 23 = (\L_2,\L_2), \quad (\L_1,\L_2) = \frac 13, 
 \quad (\a_1, \a_1) = 2.
\eeq
Now
\beq
V_{\L}\tens V_{\L_2} = V_{\L + \nu_1} \oplus V_{\L+\nu_2}\oplus V_{\L+\nu_3}
\eeq
where $\nu_i = \L_2,\L_2 - \a_2, -\L_1$ are the weights of the fundamental
representation $V_{\L_2}$. If $N_1 =0$, then the last summand does not occur.
This implies that 
\beq
X = \frac 12 (c_2(\L + \nu) - c_2(\L) - c_2(\L_2))
  = (\nu, \L +\rho) - (\L_2,\rho)
\eeq
on $V_{\L+\nu}$, which gives
\beq
X = (\frac{2N}3 + \frac n3\;,
   \;-\frac N3 + \frac n3  -1,
   \; -\frac N3 - \frac{2n}3 - 2\;)
\eeq
on $V_{\L+\nu_i}$ for $i=1,2,3$ and $\L = N \L_2 + n \L_1$.
Hence the characteristic equation of $X$ is 
\beq
(X-\frac{2N}3 -\frac n3)(X+\frac N3-\frac n3 +1)(X+\frac N3+\frac{2n}3+2) =0
\label{char-X}
\eeq
as long as $n>0$, while for $n=0$ the last factor disappears and
\beq
(X-\frac{2N}3)(X+\frac N3+1) =0.
\label{chareq-fuzzy}
\eeq

\section*{Appendix C: rewriting the Yang-Mills action}
\addcontentsline{toc}{subsection}{Appendix C: rewriting the Yang-Mills action}

The following identity holds for $su(3)$ (see e.g. \cite{fuchs-schweigert})
\beq
\sum_e d_{abe} d_{cde} = \frac 13 \(4(\d_{ac} \d_{bd} + \d_{bc} \d_{ad} 
        - \d_{ab} \d_{cd}) 
 + \sum_e (f_{ace} f_{bde} + f_{ade} f_{bce})\).
\eeq
Using the Jacobi-identity
\beq
\sum_e f_{ade} f_{bce} + f_{cae} f_{bde} + f_{abe} f_{cde} =0
\eeq
we get
\beq
\sum_e d_{abe} d_{cde} = \frac 13 \(4(\d_{ac} \d_{bd} + \d_{bc} \d_{ad} 
        - \d_{ab} \d_{cd}) 
 + \sum_e (2 f_{ade} f_{bce} + f_{abe} f_{cde} )\).
\label{dd-rewritten}
\eeq
Contracting this with $C_a C_b C_c C_d$ gives
\beq
Tr \(-(dCC)(dCC) + \frac 13 \(4 \d_{ac} \d_{bd}   
 + \sum_e (2  (fCC)_e C_d C_a f_{ade} + (fCC)(fCC)\)\) =0
\eeq
hence
\beq
Tr \(\frac 12[C,C]^2 + (C\cdot C) (C\cdot C)\) 
= Tr \(\frac 34 (dCC)(dCC) + \frac 14\sum_e (fCC) (fCC)\) 
\label{YM-rewritten}
\eeq
where $\frac 12 [C,C]^2  = C_a C_b C_a C_b - (C\cdot C) (C\cdot C)$.

\section*{Appendix D: Covariant operators for instantons}
\addcontentsline{toc}{subsection}{Appendix D: 
Covariant operators for instantons}

We must learn to work with the $\zeta = \zeta_a^{(m)}$ operators, which 
act on $V_\L$ with $\L = M \L_2 + \L_1$. To handle them, it is useful to
consider 
\beq
\tilde D:= \tilde D^{(m)} = (\frac{-M+1}6 + X)^2 - (\frac{M+1}2)^2
   = \frac{M+2}3 + \frac 12 \tilde D^{(m)}_a \tau^a
\eeq
where $X = \zeta_a \tau^a$ and \eq{D1m-def}
$$
\tilde D^{(m)}_c = {d^{ab}}_c \zeta_a \zeta_b - \frac 13(2M +7) \zeta_c.
$$
Using \eq{char-X}, it follows that $\tilde D$ 
satisfies the equation
\beq
\tilde D X = -\frac{M+8}3 \;\tilde D.
\label{DX-eq}
\eeq
It is a projector on
$V_{\L-\L_1}$, and vanishes on $V_{\L+\L_2}$ and
$V_{\L+\L_2-\a_2}$. 
This implies
\beq
\tilde D_a \zeta_a  = \zeta_a \tilde D_a = tr(\tilde D X) 
=  - \frac {M+8}3\; tr(\tilde D) =
-\frac {(M+2)(M+8)}{3}.
\label{Dt-eq}
\eeq
Similarly, from considering $tr(\tilde D \tilde D)$ one obtains
\beq
\tilde D_a \tilde D_a = \frac 13 (M+2)(10M+32),
\label{DD-eq}
\eeq
which shows that $\tilde D_a$ is an operator of order $N$.
We also note the Casimir
\beq
\zeta_a \zeta_a = \frac 12(\L,\L+2\rho) = \frac{(M+2)^2}3.
\label{tt-eq}
\eeq
Now consider $i f^{ab}_c \tilde D_a \zeta_b$. It is a covariant operator
on $V_\L$ in the sense of \eq{D-covariance}. 
However by a suitable generalization of the 
Wigner-Eckart theorem\footnote{i.e. by decomposing the intertwiner space
$V_\L \tens V_\L^*$} there are only two such
operators, which must be $\zeta_a$ and $\tilde D_a$. Therefore 
\beq
i f^{ab}_c \tilde D_a \zeta_b = \a \zeta_c + \b \tilde D_c
\eeq
for some constants $\a,\b$.
Contracting with $\zeta_c$, we get
$$
i f^{abc} \tilde D_a \zeta_b \zeta_c  = \a \zeta_c \zeta_c + \b \tilde D_c \zeta_c  = -3 \tilde D_a \zeta_a =  (M+2)(M+8).
$$
Therefore
$$
(3+\b)(M+8) = \a (M+2).
$$
Similarly,
\beq
i f^{abc} \tilde D_a [\zeta_b, \tilde D_c] = \frac i2 i f^{abc} f_{bc}^d \tilde D_a \tilde D_d
 = -6 \tilde D_a \tilde D_a = 2(\a  \zeta_c + \b \tilde D_c) \tilde D_c 
\label{fDtD}
\eeq
using \eq{ff-id} implies
$$
(3+\b) (10M+32) =  \a (M+8).
$$
Solving these gives
\beq \fbox{$
i f^{ab}_c \tilde D_a \zeta_b = -3 \tilde D_c = i f^{ab}_c  \zeta_a \tilde D_b $}
\eeq
since $\zeta_a$ is selfadjoint.

Now consider again $\tilde D X = -\frac {M+8}3\; \tilde D$. Writing it out in
components gives 
\beq
\frac 14(if+d)_{abc} \tilde D_a \zeta_b  + \frac{M+2}3 \zeta_c =
-\frac{M+8}6\; \tilde D_c 
\eeq
which using the above result gives
$$\fbox{$
d^{ab}_c \tilde D_a \zeta_b = -\frac 43(2 + M) \zeta_c -\frac 13(2M+7) \tilde D_c = d^{ab}_c \zeta_a \tilde D_b$}
$$
Next, consider
\beq
d^{ab}_c \tilde D_a \tilde D_b = \a' \zeta_c + \b' \tilde D_c.
\eeq
Contracting with $\zeta_c$ and using the previous results gives
$$
(4(2 + M)+3\b')(M+8)  -(2M+7) (10M+32)   = \a' 3(M+2)
$$
To proceed, we need some extra information on $\tilde D_a$. We claim that
on the ``bottom'' of the \rep $V_\L$ (see figure \ref{fig:inst} for
illustration), i.e. at the lowest eigenvalue 
$-\frac 1{2\sqrt{3}}(M+2)$ of $\zeta_8$, we have
$\tilde D_{1,2,3} =0$ and
$\tilde D_8 = \frac 2{\sqrt{3}} (M+ 2)$. One way to see this is 
to note that $\tilde D_{1,2,3} = x \zeta_{1,2,3}$ there by covariance, since
the multiplicity is 1 on the ``bottom''; then 
\eq{Dt-eq},\eq{DD-eq},\eq{tt-eq} together imply $x=0$. 
It can also be seen using the projector
property of $\tilde D$. The eigenvalue of $\tilde D_8$ can then be calculated
explicitly using the quadratic Casimir:
\beq
\tilde D_8 = {d^{ab}}_8 \zeta_a \zeta_b - \a \zeta_8 
 = \frac 1{\sqrt{3}}\((\sum_a \zeta_a \zeta_a) - 3(\zeta_1 \zeta_1 + \zeta_2 \zeta_2 + \zeta_3 \zeta_3)
      + \zeta_8 \zeta_8 \) - \frac 13(2M +7) \zeta_8,
\eeq
which gives $\tilde D_8 = \frac 2{\sqrt{3}} (M+ 2)$ on the bottom of  $V_\L$.
We can now calculate
\beqa
d_{8ab} \tilde D_a \tilde D_b &=& d_{8aa} \tilde D_a \tilde D_a 
 = \frac 1{\sqrt{3}}\((\sum_a \tilde D_a \tilde D_a) - 3(\tilde D_1 \tilde D_1 + \tilde D_2 \tilde D_2 + \tilde D_3 \tilde D_3)
      + \tilde D_8 \tilde D_8 \) \nn\\
 &=& \frac 1{\sqrt{3}}\((\sum_a \tilde D_a \tilde D_a) + \tilde D_8 \tilde D_8 \) \nn\\
 &=&  \frac 2{3\sqrt{3}}(M+2)(7M+20)
  =  (M +2)( -\a'\frac 1{2\sqrt{3}}  + \b' \frac 2{\sqrt{3}})
\eeqa
Therefore
$$
2(7M+20)  =  -\frac 3{2} \a' + 6\b'.
$$
Together with the above this gives 
$\a'=-4(M+2),\; \b' = \frac 23(2M+7)$ and
$$\fbox{$
d^{ab}_c \tilde D_a \tilde D_b = -4(M+2) \zeta_c + \frac 23(2M+7) \tilde D_c    $}
$$
Together with
$\tilde D^2 = (2M+6)\tilde D$
written in components this implies
$$\fbox{$
i f^{ab}_c \tilde D_a \tilde D_b = 4(M+2) \zeta_c + 2(2M+7) \tilde D_c $}
$$

\section*{Appendix E: Explicit form of the instanton field}
\addcontentsline{toc}{subsection}{Appendix E: Explicit form of the
  instanton field}
\label{sec:inst-explicit}

We first calculate the  field 
\beq
A_a'(x) = \a' \zeta^{(m)}_a - \xi_a 
\eeq
for the above configuration of  $U(2)$ gauge theory explicitly.
We have to choose a gauge, i.e. a nice embedding of 
$V_{N \L_2} \oplus V_{N \L_2} \subset V_{M\L_2+\L_1}$. 
More precisely, we must find a suitable basis for these two spaces
to define the embedding of $\xi_a \tens \one$, and calculate the difference.
Again we use the 
Gelfand-Tsetlin basis for $V_{M\L_2+\L_1}$ and $V_{N \L_2} \oplus V_{N \L_2}$
(see Appendix F), where the operators can be calculated
explicitly.
One finds \eq{instanton-reps-1}
\beqa
2 \sqrt{3} \zeta^{(m)}_{8}   &=& 2\sqrt{3}\xi_{8}  + (-2m+1), \quad
  2 \zeta_{3}^{(m)}  = 2 \xi_3 + \sigma^3, \nn\\
(\zeta_4 \pm i \zeta_5)^{(m)} &=& (\xi_4 \pm i\xi_{5})^{(M+\e)} , \nn\\
(\zeta_6 - i \zeta_7)^{(m)} &=& (\xi_6 - i\xi_{7})^{(M+\e)}
        (1 -h(x)\frac{\s^3}{2M})  + \frac{x_4 - ix_{5}}{\sqrt{3}}\;
      h(x) \frac{\sqrt{3}}{\sqrt{1+2x_8}}
       \frac 12 \s_+ \nn\\
  &=&  (\xi_6 - i\xi_{7})^{(M+\e)}
      - \frac{x_6 - ix_{7}}{\sqrt{3}}\; h(x) \frac 12 \s_3
      + \frac{x_4 - ix_{5}}{\sqrt{3}}\; h(x) \frac{\sqrt{3}}{\sqrt{1+2x_8}}
       \frac 12 \s_+ \nn\\
(\zeta_6 + i \zeta_7)^{(m)} &=& (\xi_6 + i\xi_{7})^{(M+\e)} 
    (1 -h(x)\frac{\s^3}{2M})   +  \frac{x_4 + ix_{5}}{\sqrt{3}}\; 
   h(x) \frac{\sqrt{3}}{\sqrt{1+2x_8}}
    \frac 12 \s_- \nn\\
 &=& (\xi_6 + i\xi_{7})^{(M+\e)}
      - \frac{x_6 + ix_{7}}{\sqrt{3}} h(x) \frac 12 \s_3
   + \frac{x_4 + ix_{5}}{\sqrt{3}}\; h(x) \frac{\sqrt{3}}{\sqrt{1+2x_8}}
    \frac 12 \s_- \nn\\
(\zeta_1 \pm i \zeta_2)^{(m)} &=& (1-h(x)\frac{\s^3}{2M})(\xi_1 \pm i\xi_{2}) +
  h(x) \frac{\sqrt{1+2x_8}}{\sqrt{3}} \frac 12 \s^\pm  \nn\\
  &=& (\xi_1 \pm i\xi_{2}) 
  -  \frac{x_1 + ix_{2}}{\sqrt{3}} h(x) \frac 12 \s_3
 + h(x) \frac{\sqrt{1+2x_8}}{\sqrt{3}} \frac 12 \s^\pm  
\label{zeta-explicit}
\eeqa
where \eq{hvonx}
$$
h(x) =  \frac{3}{2+x_8-\sqrt{3} x_3}
$$
and $\s_\pm = \s_1\pm i\s_2$.
Using $\a' = 1+\frac{2m-1}{2N} +o(1/N^2)$, this gives
\beqa
A_8' &=&  \frac {2m-1}2\; A_8^{mono}, \quad
A_3' =    \frac {2m-1}2\; A_3^{mono} +  \frac 1{2} \sigma^3, \nn\\
A_4' \pm i A_5' &=&  \frac {2m-1}2\; (A_4^{mono} \pm i A_5^{mono})
  + \frac{\sqrt{3}}{2}\frac{x_4\pm i x_5}{2x_8+1} \frac 1{2} \sigma^3 \nn\\
A_6' \pm i A_7' &=&  \frac {2m-1}2\; (A_6^{mono} \pm i A_7^{mono})      
   + \frac{\sqrt{3}}{2}\frac{x_6\pm i x_7}{2x_8+1} \frac 1{2}\sigma^3 \nn\\
 && - \frac{x_6 \pm ix_{7}}{\sqrt{3}}\; h(x) \frac 12 \s_3
      + \frac{x_4 \pm ix_{5}}{\sqrt{3}}\; h(x) \frac{\sqrt{3}}{\sqrt{1+2x_8}}
       \frac 12 \s_\mp \nn\\
A_1' \pm i A_2' &=&  \frac {2m-1}2\; (A_1^{mono} \pm i A_2^{mono})
 - h(x) \frac{x_1\pm i x_2 }{\sqrt{3} }  \frac 12 \s_3
    +  h(x) \frac{\sqrt{1+2x_8}}{\sqrt{3}}
       \frac 12 \s^\pm \nn\\
\label{instantonA-1}
\eeqa
One can check using \eq{constraint-explicit} that $A_a x_a =0$, as it must be.
At first sight, the $U(1)$ part looks like a half-integer monopole,
which is not allowed. However, this can be rewritten as
\beqa
A_8' &=& A_8^{mono} Q 
    + \frac 1{\sqrt{3} } (x_8-1) \frac 12 \s_3  \nn\\
A_3' &=& A_3^{mono} Q  
   +  \frac 1{\sqrt{3}} (x_3+\sqrt{3}) \frac 12 \sigma^3, \nn\\
A_4' \pm i A_5' &=& (A_4^{mono} \pm i A_5^{mono}) Q
  + \frac1{\sqrt{3}} (x_4\pm i x_5) \frac 1{2} \sigma^3 \nn\\
A_6' \pm i A_7' &=& (A_6^{mono} \pm i A_7^{mono}) Q 
    + \frac{x_6\pm i x_7}{\sqrt{3}} (1-h(x)) \frac 12 \s_3
    +  h(x) \frac{x_4\pm i x_5}{\sqrt{1+2x_8}}
\frac 12 \s_\mp \nn\\
A_1' \pm i A_2' &=&  (A_1^{mono} \pm i A_2^{mono}) Q 
    + \frac{x_1\pm i x_2}{\sqrt{3}} (1-h(x)) \frac 12 \s_3
       + h(x) \frac{\sqrt{1+2x_8}}{\sqrt{3}}
         \frac 12 \s^\pm   \nn
\eeqa
where the $U(1)$ monopole part is associated to the generator 
\beq
Q = \frac {2m-1}2 -\frac 12 \s_3 = \(\begin{array}{cc} m-1 & 0 \\ 
                                                    0 & m\end{array}\)
\label{Q}
\eeq
which has ``charge 1'',
i.e. $e^{2\pi i Q} =1$ but $e^{\pi i Q} \neq 1$.
This is consistent with the usual quantization condition.
We therefore write
\beq 
A_a' =  Q A_a^{mono}  + (A_a^{inst})'.
\eeq 
Projecting out the non-tangential part as in \eq{instanton-proj}
is straightforward in the commutative  limit, writing
$A_a = A_a' + (\frac 1{\sqrt{3}} d_{abc} x_b A_c' -\frac 13 A'_a)$.
Since this map is linear and the monopole part is already tangential,
we have
$$
A_a = Q A_a^{mono} +  A_a^{inst} =  
 Q A_a^{mono}  
+ (A_a^{inst})' + (\frac 1{\sqrt{3}} d_{abc} x_b (A_c^{inst})'
     -\frac 13 (A_a^{inst})').
$$
Explicitly, this is is using \eq{constraint-explicit}
\beqa
&& A_1^{inst}\pm i A_2^{inst} =
  \frac{x_1\pm i x_2}{\sqrt{3}} (2-h(x))\; \frac 12 \s_3  
 + \sqrt{\frac{1+2 x_8}3} (h(x)-1) \frac 12 \s_\pm  \nn\\
&&A_3^{inst} = (2\frac{x_3}{\sqrt{3}} +1-\frac{1+2x_8}3 h(x)) \frac 12 \s_3
  - \frac{\sqrt{2x_8+1}}3 h(x) \frac 12 (x_1 \s_1 + x_2 \s_2) \nn\\
&& A_4^{inst} = \frac {h(x)}{\sqrt{1+2x_8}^3}\( x_4 x_5 (x_7 \s_1 + x_6 \s_2)
       + \frac 16 (3 x_4^2 -3 x_5^2 -(1+2x_8)^2) (x_6 \s_1 - x_7 \s_2)\) \nn\\
  &&\quad  - \frac{h(x)}{\sqrt{3}(1+2x_8)} (x_6^2+x_7^2) x_4 \s_3  \nn\\
&&A_5^{inst} = \frac {h(x)}{\sqrt{1+2x_8}^3}\( x_4 x_5 (-x_6 \s_1 - x_7 \s_2)
       + \frac 16 (3 x_4^2 -3 x_5^2 -(1+2x_8)^2) (x_7 \s_1 - x_6 \s_2)\) \nn\\
  &&\quad  - \frac{h(x)}{\sqrt{3}(1+2x_8)} (x_6^2+x_7^2) x_5 \s_3  \nn\\
&& A_6^{inst} \pm i A_7^{inst} = \textstyle
 \frac {h(x)}{2\sqrt{1+2x_8}^3}(x_6^2+x_7^2) (x_4\pm i x_5) (\s_1 \mp i\s_2)
  + \frac 1{2\sqrt{3}} (2-h(x)) (x_6 \pm i x_7) \s_3   \nn\\
&& A_8^{inst} = \textstyle \(- \frac 13 x_3 +\frac 1{6\sqrt{3}} 
      (-2+(h(x)-1)(2x_1^2+2x_2^2-x_6^2-x_7^2)
    -2 x_3^2 + x_4^2+x_5^2 +2x_8^2)\) \s_3 \nn\\
  && \textstyle\quad + \frac{h(x)}{6\sqrt{1+2x_8}}\((x_4 x_6 + x_5 x_7
            -\frac 2{\sqrt{3}} (1+2x_8) x_1) \s_1
       + (x_5 x_6 - x_4 x_7
            -\frac 2{\sqrt{3}} (1+2x_8) x_1) \s_2\)\nn
\eeqa
which can be seen to be tangential.

\section*{Appendix F: Gelfand-Tsetlin basis for monopoles and
  instantons}
\addcontentsline{toc}{subsection}{Appendix F: Gelfand-Tsetlin basis 
for monopoles and instantons}
\label{sec:gelftset}

We need the explicit form for the action of the Lie algebra
on $V_{N\L_2 +n \L_1}$. This is given in terms of the 
Gelfand-Tsetlin basis (see \cite{molev} for a review), which works as follows.
Consider complexified 
$gl(3) \cong su(3) \oplus u(1)$, with generators being the 
elementary matrices $E_{ij}$ and the Cartan subalgebra $\mh$ generated by
$E_{ii}$. Irreps of $gl(3)$ are highest weight reps $V_\mu$
with highest weights 
being labeled by 3 
complex numbers $\mu=(\mu_1,\mu_2,\mu_3)$ 
such that
\beq
\mu_1 \geq \mu_2\geq \mu_3.
\eeq
The  highest weight vector satisfies
$
E_{ii}\;|\mu\rangle=\mu_i |\mu\rangle
$
for $i=1,2,3$, and
$
E_{ij}\;|\mu\rangle=0
$
for $1\leq i<j\leq 3$. 
The relation with the $su(3)$ weights is as follows:
\beq
\L_1 \cong (1,0,0), \qquad \L_2 \cong (1,1,0)
\eeq
where the overall sum is the $u(1)$ charge and is ignored for $su(3)$.
The relations among the generators are
\beqa
H_{\a_1} &=& E_{11} - E_{22} = H_3, \qquad H_{\a_2} = E_{22} - E_{33},\nn\\
\sqrt{3}\; H_8 &=& H_{\a_1} + 2 H_{\a_2} = E_{11} + E_{22}-2 E_{33}, \nn\\
X_1^+ &=& E_{1,2}, \qquad X_2^+ = E_{2,3}, \nn\\
X_1^- &=& E_{2,1}, \qquad X_2^- = E_{3,2}, \nn
\eeqa
One then associates to such a weight $\mu$ the pattern
\beq
\cP = \(\begin{array}{c}
\mu_{1}\quad\mu_{2}\quad\mu_{3}\\
  p \quad q\\
 r \end{array} \) = \(\begin{array}{c}
\mu_{1,3}\quad\mu_{2,3}\quad\mu_{3,3}\\
  \mu_{1,2}\quad\mu_{2,2} \\
   \mu_{1,1}\end{array} \)
\eeq
where
\beqa
\mu^{}_1 &\geq& p \geq \mu^{}_{2} \geq q\geq\mu^{}_{3}, \nn\\
p &\geq&  r \geq  q \nn
\eeqa
One can then show that a basis of $V_\mu$ is given by the 
orthonormal weight vectors
$$
|\cP \rangle \equiv |\mu ; p,q, r \rangle \equiv  |\cP_\mu \rangle,
$$ 
and the action of the $gl(3)$ generators
is 
\beqa
\label{GT}
E_{1,1}\; |\cP\rangle &=& r\; |\cP\rangle, \nn\\
E_{2,2}\; |\cP\rangle &=& (p+q-r)\; |\cP\rangle, \nn\\
E_{3,3}\; |\cP\rangle &=& (\mu_1+\mu_2+\mu_3-p-q)\; |\cP\rangle, \nn\\
H_{\a_1}    \; |\cP\rangle &=& (2r-p-q) \; |\cP\rangle, \nn\\
H_{\a_2}    \; |\cP\rangle &=& (2(p+q)-r-(\mu_1+\mu_2+\mu_3)) \; |\cP\rangle, \nn\\
E_{1,2}\; |\cP\rangle &=& A^1_1(\cP)\; |\cP_{r+1}\rangle,  \nn\\
E_{2,1}\; |\cP\rangle &=& A^1_1(\cP_{r-1}) \;|\cP_{r-1}\rangle,  \nn\\
E_{2,3}\; |\cP\rangle &=& A^1_2(\cP) \;|\cP_{p+1}\rangle 
                      + A^2_2(\cP) \;|\cP_{q+1}\rangle,\nn\\
E_{3,2}\; |\cP\rangle &=& A^1_2(\cP_{p-1}) \;|\cP_{p-1}\rangle 
                      + A^2_2(\cP_{q-1}) \;|\cP_{q-1}\rangle\nn
\eeqa
where $|\cP_{p-1}\rangle$ means that $p$ is replaced by $p-1$, etc,
and it is supposed
that $|\cP\rangle =0$ if $\cP$ is not a pattern.
Here
\beqa
A^1_1(\cP) &=& \(-(\mu_{1,2} - \mu_{1,1})(\mu_{2,2} -\mu_{1,1}-1)\)^{1/2}, 
               \nn\\
A^1_2(\cP) &=&  \(-\frac{(\mu_{1,3}-\mu_{1,2})(\mu_{2,3} -\mu_{1,2}-1)
                                          (\mu_{3,3}-\mu_{1,2}-2)
                         (\mu_{1,1} - \mu_{1,2}-1)}
            {(\mu_{2,2}-\mu_{1,2}-1)(\mu_{2,2}-\mu_{1,2}-2)}\)^{1/2}, \nn\\
A^2_2(\cP) &=&  \(-\frac{(\mu_{1,3}-\mu_{2,2}+1)(\mu_{2,3} -\mu_{2,2})
                                        (\mu_{3,3}-\mu_{2,2}-1)
                       (\mu_{1,1} - \mu_{2,2})}
                     {(\mu_{1,2}-\mu_{2,2}+1)(\mu_{1,2}-\mu_{2,2})}\)^{1/2}.
\eeqa
Consider first $N \L_2 \cong (N,N,0)$. The associated pattern is 
\beq
\cP = \(\begin{array}{c}
\mu_{1}\quad\mu_{2}\quad\mu_{3}\\
  p \quad q\\
 r \end{array} \) = \(\begin{array}{c}
\mu_{1,3}\quad\mu_{2,3}\quad\mu_{3,3}\\
  \mu_{1,2}\quad\mu_{2,2} \\
   \mu_{1,1}\end{array} \)
 = \(\begin{array}{c}
N\quad N\quad 0\\
  N \quad q\\
   r \end{array} \)
\eeq
hence $N \geq r \geq q \geq 0$,
and the highest weight state is 
$q=r=N$.
Then
\beqa
A^1_1(\cP) &=& ((N - r)(r+1-q))^{1/2},  \nn\\
A^1_2(\cP) &=&   0, \nn\\
A^2_2(\cP) &=& ((q+1)(r - q))^{1/2}.
\eeqa
Hence
\beqa
H_{\a_1} \; |\cP\rangle &=& (2r-q-N) \; |\cP\rangle, \nn\\
H_{\a_2} \; |\cP\rangle &=& (2q-r) \; |\cP\rangle, \nn\\
\sqrt{3} H_{8} \; |\cP\rangle &=& (3q-N) \; |\cP\rangle  
          = H_{\a_1} + 2H_{\a_2} \; |\cP\rangle , \nn\\
X_1^+ \; |\cP\rangle  &=& \sqrt{(N - r)(r+1-q)} |\cP_{r+1}\rangle,  \nn\\
X_1^-\; |\cP\rangle &=& \sqrt{(N - r+1)(r-q)} \;|\cP_{r-1}\rangle,  \nn\\
X_2^+\; |\cP\rangle &=& \sqrt{(q+1)(r - q)} \;|\cP_{q+1}\rangle,\nn\\
X_2^-\; |\cP\rangle &=& \sqrt{q(r - q+1)} \;|\cP_{q-1}\rangle\nn\\
X_\theta^+\; |\cP\rangle &=& -\sqrt{(q+1)(N-r)}\;|\cP_{r+1,q+1}\rangle\nn\\
X_\theta^-\; |\cP\rangle &=& -\sqrt{q(N-r+1)}\;|\cP_{r-1,q-1}\rangle\nn
\eeqa
using $X_\theta^+ = [X_1^+,X_2^+]$.

It turns out that this basis is not useful to calculate the
instanton fields (it would lead to a singularity at the north pole). 
A better basis is obtained by applying a Weyl rotation
$S_2$ along the root $\a_2$,
defining
\beq
|\cP\rangle' = S_2 |\cP\rangle
\eeq
The corresponding automorphism $T_2(X) = S_2 X S_2^{-1}$ is
\beqa
T_2(H_2) &=& -H_2, \qquad T_2(H_1) = H_1 + H_2 \nn\\
T_2(X^\pm_2) &=& - X^\mp_2, \qquad
T_2(X_1^\pm) = -X_\theta^\pm, \qquad
T_2(X_\theta^\pm) = X_1^\pm.\nn
\eeqa
Hence 
\beq \begin{array}{rll}
H_{\a_1}\; |\cP\rangle' =& (r+q-N) \; |\cP\rangle'
  &=  (-a+b)  \; |\cP\rangle', \nn\\
\sqrt{3}\;H_{8} \; |\cP\rangle' =& (3r-3q-N) \; |\cP\rangle'
  &= (2N-3(a+b)) \; |\cP\rangle', \nn\\
X_\theta^+ \; |\cP\rangle' =& -\sqrt{(N -r)(r+1-q)}|\cP_{r+1}\rangle'
  &=  -\sqrt{a(N-a-b+1)}\;|\cP_{a-1}\rangle', \nn\\
X_\theta^-\; |\cP\rangle' =& -\sqrt{(N - r+1)(r-q)}\;|\cP_{r-1}\rangle'
  &= -\sqrt{(a+1)(N-a-b)}\;|\cP_{a+1}\rangle', \nn\\
X_2^-\; |\cP\rangle' =& -\sqrt{(q+1)(r - q)} \;|\cP_{q+1}\rangle'
  &=  -\sqrt{(b+1)(N-a-b)} \;|\cP_{b+1}\rangle',  \nn\\
X_2^+\; |\cP\rangle' =& -\sqrt{q(r - q+1)} \;|\cP_{q-1}\rangle'
  &= -\sqrt{b(N-a-b+1)} \;|\cP_{b-1}\rangle',  \nn\\
X_1^+\; |\cP\rangle' =& -\sqrt{(q+1)(N-r)}\;|\cP_{r+1,q+1}\rangle'
  &=  -\sqrt{a(b+1)} |\cP_{a-1,b+1}\rangle',  \nn\\
X_1^-\; |\cP\rangle' =& -\sqrt{q(N-r+1)}\;|\cP_{r-1,q-1}\rangle'
  &= -\sqrt{(a+1)b} \;|\cP_{a+1,b-1}\rangle'\nn
\end{array}
\label{GT-generators}
\eeq
introducing the parameters $a=N-r, b=q$ 
which range from $0 \leq a,b \leq N, 0 \leq a+b \leq N$.
The north pole (with maximal eigenvalue of $H_8$) 
satisfies $r=N,q=0$ and $\sqrt{3} H_8 = 2N$, or
$a=b=0$. The parameters $a,b$ measure the deviation from the north pole.

To find the monopole gauge field, we have to repeat the same
calculation replacing $N$ by $M=N-m$. Since we want to match the north
poles (i.e. $A=0$ at the north pole), we can simply match the $a$ and
$b$ parameters. This defines a map $V_{M \L_2} \mapsto V_{N \L_2}$
except at the south sphere, and the generators of  $V_{M \L_2}$
can be expressed  in terms of the generators of $V_{N \L_2}$ as
\beqa
H_{\a_1}^{(M)} &=& H_{\a_1}, \nn\\
\sqrt{3} H_{8}^{(M)} &=& \sqrt{3} H_{8} +2(M-N), \nn\\
{X_1^\pm}^{(M)} &=& X_1^\pm, \nn\\
{X^\pm_2}^{(M)} &=& 
   X^\pm_2(1+\frac{3}2\frac{M-N}{\sqrt{3} H_8+N} + o(1/N^2)), \nn\\
{X^\pm_\theta}^{(M)} &=&
   X^\pm_\theta(1+\frac{3}2\frac{M-N}{\sqrt{3} H_8+N}+o(1/N^2)).
\label{monopole-gens}
\eeqa
Using
\beq
\xi_8 = T_8 =\frac 12 H_8, \quad \xi_3 = T_{3} = \frac 12 H_3 \quad
\xi_1 \pm i \xi_2 = X^\pm_1, \quad
\xi_4 \pm i \xi_5 = X^\pm_\theta, 
\quad \xi_6 \pm i \xi_7 = X^\pm_2, 
\label{xi-X-matching}
\eeq
this leads to \eq{monopole-generators-explicit}.

\subsection*{Instantons}

Next consider $N \L_2 +n \L_1\cong (n+N,N,0)$.
The associated pattern is 
\beq
\cP = \(\begin{array}{c}
\mu_{1}\quad\mu_{2}\quad\mu_{3}\\
  p \quad q\\
 r \end{array} \) = \(\begin{array}{c}
\mu_{1,3}\quad\mu_{2,3}\quad\mu_{3,3}\\
  \mu_{1,2}\quad\mu_{2,2} \\
   \mu_{1,1}\end{array} \)
 = \(\begin{array}{c}
N+n\quad N\quad 0\\
 \quad N + \e \qquad q\\
   r \end{array} \)
\eeq
where $n\geq \e \geq 0$ and  $N+\e\geq r \geq q \geq 0$ and $N \geq q$. 
We only need $n=1$, so that $\e = 0,1$. Then 
the highest weight state is given by 
$\e=1, q=N, r=N+1$.
Furthermore $A^1_2(\cP) =0$ for $\e=1$, and 
\beqa
A^1_1(\cP) &=& \sqrt{(N+\e - r)(r+1-q)},  \nn\\
A^1_2(\cP) &=& \d_{\e,0}\; 
     \sqrt{N-r+1}\(\frac{(N+2)}{(N-q+1)(N-q+2)}\)^{1/2}, \nn\\
A^2_2(\cP) &=& \sqrt{(q+1)(r - q)}\; 
        \(\frac{(N+2-q)(N -q)}{(N+\e-q+1)(N+\e-q)}\)^{1/2}.
\eeqa
Note that since $\e=0,1$ only, this can be written as
\beq
\(\frac{(N+2-q)(N -q)}{(N+\e-q+1)(N+\e-q)}\)^{1/2}
 =  \sqrt{1-\frac{\s_3}{N+1-q}}
\eeq
where $\s_3 = 2\e-1$.
We also define
\beq
h(q) = \sqrt{\frac{N(N+2)}{(N-q+1)(N-q+2)}}.
\eeq
Then 
\beqa
H_{\a_1} \; |\cP\rangle &=& (2r-q-\e-N) \; |\cP\rangle, \nn\\
H_{\a_2} \; |\cP\rangle &=& (2q+2\e-r-1) \; |\cP\rangle, \nn\\
\sqrt{3} H_8\; |\cP\rangle &=& (3q+3\e-2-N) \; |\cP\rangle, \nn\\
X_1^+\; |\cP\rangle &=& \sqrt{(N+\e - r)(r+1-q)}\; |\cP_{r+1}\rangle,  \nn\\
X_1^-\; |\cP\rangle &=& \sqrt{(N+\e - r+1)(r-q)} \;|\cP_{r-1}\rangle,  \nn\\
X_2^+\; |\cP\rangle &=& \d_{\e,0}\; \sqrt{(N-r+1)/N}h(q) 
             |\cP_{p+1}\rangle\nn\\ 
 && + \sqrt{(q+1)(r - q)} \sqrt{1-\frac{\s_3}{N+1-q}}\;|\cP_{q+1}\rangle,\nn\\
X_2^-\; |\cP\rangle &=& \d_{\e,1}\;\sqrt{(N-r+1)/N}h(q)\;|\cP_{p-1}\rangle\nn\\
 &&   + \sqrt{q(r -q +1)} \sqrt{1-\frac{\s_3}{N+2-q}}\;|\cP_{q-1}\rangle\nn\\
X_\theta^+\; |\cP\rangle &=& \d_{\e,0}\; 
   \sqrt{(r+1-q)/N}h(q) |\cP_{r+1,p+1}\rangle\nn\\ 
 && - \sqrt{(q+1)(N+\e-r)}\sqrt{1-\frac{\s_3}{N+1-q}}|\cP_{r+1,q+1}\rangle\nn\\
X_\theta^-\; |\cP\rangle &=& \d_{\e,1}\; 
   \sqrt{(r-q)/N} h(q) |\cP_{r-1,p-1}\rangle\nn\\ 
 && - \sqrt{q(N+\e-r+1)}\sqrt{1-\frac{\s_3}{N+2-q}}|\cP_{r-1,q-1}\rangle\nn
\eeqa
We should find that this is approximately $\xi_a \tens \one_2$. 
This is not natural in this basis (the singularity would be at the
north pole). A more suitable basis is found after a Weyl 
reflection $|\cP\rangle' = S_2 |\cP\rangle$, 
as in the last section.
The north pole then satisfies $q=0$, and either $\e=0, r=N$ or
$\e=1,r=N+1$ with $\sqrt{3}H_8= N = 2N$ as it should be.

\begin{figure}[htpb]
\begin{center}
\epsfxsize=3in
  \vspace{0.3in} 
   \epsfbox{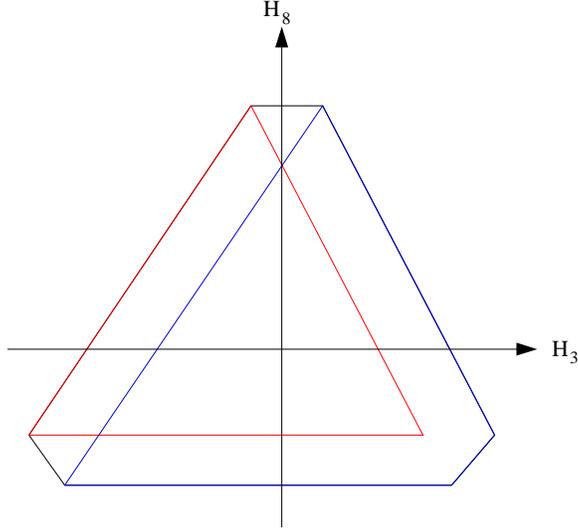}
\end{center}
 \caption{decomposition of $V_{N\L_2 + \L_1}$ into 
the eigenspaces of $\e=0, \e=1$}
\label{fig:inst}
\end{figure}

Figure \ref{fig:inst} shows schematically the decomposition 
of $V_{N\L_2 + \L_1}$ 
into the eigenspaces of $\e=0,1$. 
In terms of the parameters $a=N +\e-r, b=q$ with the range
$0 \leq a \leq N+\e$, $0\leq b \leq N$, $a+b \leq N+\e$
we get
\beqa 
H_{\a_1} \; |\cP\rangle' &=& (b-a+2\e-1) \; |\cP\rangle', \nn\\
\sqrt{3} H_8\; |\cP\rangle' &=& (2N-3(a+b)+1) \; |\cP\rangle', \nn\\
X_\theta^+\; |\cP\rangle' &=& -\sqrt{a(N-a-b+\e+1)}\;|\cP_{a-1}\rangle', \nn\\
X_\theta^-\; |\cP\rangle' &=& -\sqrt{(a+1)(N-a-b+\e)}\;|\cP_{a+1}\rangle',\nn\\
X_2^-\; |\cP\rangle' &=& \textstyle -\d_{\e,0}\;
  \sqrt{\frac{a+1}{N}} h(b) |\cP_{\e+1,a+1}\rangle'
  -\sqrt{(b+1)(N+\e-a-b)}\sqrt{1-\frac{\s_3}{N+1-b}}\;|\cP_{b+1}\rangle',\nn\\
X_2^+\; |\cP\rangle' &=& -\d_{\e,1}\;
  \sqrt{\frac aN} h(b)\; |\cP_{\e-1,a-1}\rangle'
  - \sqrt{b(N+\e-a-b +1)} \sqrt{1-\frac{\s_3}{N+2-b}}\;|\cP_{b-1}\rangle'\nn\\
X_1^+\; |\cP\rangle' &=& \d_{\e,0}\; 
 \sqrt{\frac{N-a-b+1}N} h(b) |\cP_{\e=1}\rangle'
    - \sqrt{(b+1)a}\sqrt{1-\frac{\s_3}{N+1-b}}|\cP_{a-1,b+1}\rangle'\nn\\
X_1^-\; |\cP\rangle' &=& \d_{\e,1}\; 
   \sqrt{\frac{N+1-a-b}N} h(b) |\cP_{\e=0}\rangle'
    - \sqrt{b(a+1)}\sqrt{1-\frac{\s_3}{N+2-b}}|\cP_{a+1,b-1}\rangle'\nn   
\eeqa
It is now obvious that the representation splits naturally
into two subspaces with $\e=0$ and $\e=1$, each of which can be mapped
to $V_{N \L_2}$ by matching the parameters $a,b$ with the
corresponding basis of the previous section. 
Looking at the bounds for $a$ and $b$, this works perfectly
as long as $a+b \leq N$. 
Hence we have a map
$$
V_{N \L_2+\L_1} \mapsto V_{N \L_2} \tens \C^2 = \(\begin{array}{c} \cP_{\e=1} 
                                \\ \cP_{\e=0}\end{array} \)
$$
except for the lowest eigenvalue of $H_8$, which is the south sphere.
This is the ``coordinate patch'' of $\C P^2$ where the instanton gauge
field will be well-defined.
Using this identification, we can express the above generators in
terms of the generators for $V_{N \L_2}$ by comparing with \eq{GT-generators},
i.e. by writing the maps on $V_{N \L_2} \tens \C^2$  in the form
\beq
C_a = C_{a,\a} \s^\a 
\eeq
where $\s_3 = 2\e-1$ etc. 
For large $N$, we can write
$$
1-\frac bN = \frac 13 (2+x_8-\sqrt{3} x_3), \qquad
1-\frac{a+b}N = \frac 13(1+2x_8)
$$ 
so that using \eq{xi-X-matching} and \eq{GT-generators} we have 
e.g.
\beqa
-\sqrt{(a+1)/N} h(b) |\cP_{\e+1,a+1}\rangle' &=&  
  \frac{\xi_4-i\xi_5}N \sqrt{\frac 3{1+2x_8}} h(x)\;|\cP_{\e+1,a}\rangle'\nn\\
-\sqrt{a/N} h(b) |\cP_{\e-1,a-1}\rangle' &=&  
  \frac{\xi_4+i\xi_5}N  \sqrt{\frac 3{1+2x_8}} h(x)\; |\cP_{\e+1,a}\rangle' \nn
\eeqa
where
\beq
h(x) =  \sqrt{\frac{N (N+2)}{(N-b+1)(N-b+2)}}  
  =  \frac{3}{2+x_8-\sqrt{3} x_3}
\label{hvonx}
\eeq
note also 
\beqa
\sqrt{\frac{(N-a-b+1)(N+2)}{(N-b+1)(N-b+2)}} &=& 
     \sqrt{\frac{1+2x_8}3} h(x),\nn\\
\sqrt{1-\frac{\s_3}{N+2-b}}  &=& \sqrt{1-\frac{h(x)}{N} \s_3 }.
\eeqa
Then the above operators written in this notation are as follows:
\beqa
\sqrt{3} H_{8}     &=& (2\sqrt{3} \xi_8 +1) \one, \nn\\
H_{\a_1}  &=& 2 \xi_3 + \sigma^3, \nn\\
X_\theta^+ &=& \(\begin{array}{cc} (\xi_4+i\xi_5)^{(N+1)} & 0 \\
                                0  & (\xi_4+i\xi_5)^{(N)}\end{array} \), \nn\\ 
X_\theta^- &=& \(\begin{array}{cc} (\xi_4-i\xi_5)^{(N+1)} & 0 \\
                              0  &(\xi_4-i\xi_5)^{(N)} \end{array} \), \nn\\ 
X_2^-  &=& \(\begin{array}{cc}  
       (\xi_6-i\xi_7)^{(N+1)} &  0\; \\
         0 & (\xi_6-i\xi_7)^{(N)} \end{array} \)
          \sqrt{1-\frac{h(x)}{N} \s_3 } \nn\\
   &&     + \frac{\xi_4-i\xi_5}N 
          \sqrt{\frac 3{1+2x_8}} h(x)\; \frac 12\s^+ ,\nn\\
X_2^+  &=& \(\begin{array}{cc} (\xi_6+i\xi_7)^{(N+1)} &  0\\
                   0 &  (\xi_6+i\xi_7)^{(N)} \end{array} \) 
          \sqrt{1-\frac{h(x)}{N} \s_3 }  \nn\\
  && +  \frac{\xi_4+i\xi_5}N \sqrt{\frac 3{1+2x_8}} h(x)\; \frac 12 \s^-, \nn\\
X_1^+  &=&  (\xi_1+i\xi_2)  \sqrt{1-\frac{h(x)}{N} \s_3 }
    + \sqrt{\frac{1+2x_8}3} h(x) \frac 12 \s^+, \nn\\
X_1^-  &=&  (\xi_1-i\xi_2)  \sqrt{1-\frac{h(x)}{N} \s_3 }
   + \sqrt{\frac{1+2x_8}3} h(x) \frac 12 \s^- , \nn
\label{instanton-reps-1}
\eeqa 
using   
$\frac 12 \s^+ = \(\begin{array}{cc} 0  & 1 \\0  & 0 \end{array} \)$ etc.
We can now expand
\beq
 \sqrt{1-\frac{h(x)}{N} \s_3 } = 1-\frac{h(x)}{2N} \s_3 +o(1/N^2)
\eeq
which  for large $N$ leads to \eq{zeta-explicit}.

Note that the ``south sphere'' is given by $a+b = max = N+1$ and
belongs to $\e=1$.

\end{document}